\documentclass[useAMS,usenatbib]{mn2e}

\usepackage{amsmath}
\usepackage{amssymb}
\usepackage[dvipsnames]{xcolor}
\usepackage{graphicx}
\usepackage{graphics}
\usepackage{hyperref}
\hypersetup{
     colorlinks   = true,
     citecolor    = blue
}

\def\apj{{ ApJ}}
\def\apjl{{ApJL}}
\def\apjs{{ ApJS}}

\def\mnras{{ MNRAS}}
\def\araa{{ ARA\&A}}

\def\nat {{ Nature}}

\def\ssr{{ Space Sci. Rev.}}

\def\prd{{ Phys. Rev. D}}

\def\mr{\mathrm}

\def\msun{M_{\rm \odot}}
\def\mstar{M_*}
\def\mBH{M_{\rm BH}}
\def\mBHmax{M_{\rm BH,\max}}
\def\rg{r_{\rm g}}
\def\rp{r_{\rm p}}

\def\rinit{r_{\rm init}}
\def\thetainit{\theta_{\rm init}}
\def\phiinit{\phi_{\rm init}}
\def\Rsun{R_\odot}

\def\Rhm{R_{\rm hm}}

\def\Lcapt{L_{\rm capt}}
\def\LTD{L_{\rm TD}}
\def\Llc{L_{\rm lc}}
\def\GammaTDE{\Gamma_{\rm TDE}}
\def\Gammacapt{\Gamma_{\rm capt}}
\def\Gammalc{\Gamma_{\rm lc}}
\def\fTDE{f_{\rm TDE}}

\def\tage{t_{\rm age}}
\def\Gyr{{\rm Gyr}}

\def\VRL{V_{\rm RL}}
\def\tr{\Tilde{r}}
\def\ttheta{\Tilde{\theta}}
\def\tphi{\Tilde{\phi}}

\newcommand{\lrb}[1]{\left({#1}\right)}
\newcommand{\lrsb}[1]{\left[{#1}\right]}
\newcommand{\lara}[1]{\left\langle{#1}\right\rangle}
\newcommand{\abs}[1]{\left|#1\right|}

\newcommand{\myemail}{haotse813@gmail.com}

\title[TDEs by spinning BHs]{Tidal disruption rate suppression by the event horizon of spinning black holes}
\author[Huang \& Lu]
  {Hao-Tse Huang$^{1,2}$\thanks{\myemail} and Wenbin Lu$^{1}$\thanks{wenbinlu@berkeley.edu}\\
    $^1$Departments of Astronomy and Theoretical Astrophysics Center, UC Berkeley, Berkeley, CA 94720, USA\\
    $^2$Department of Physics, The Chinese University of Hong Kong, Shatin, N.T., Hong Kong S.A.R., China\\
    }

\begin{document}
\label{firstpage}
\maketitle

\begin{abstract}
The rate of observable tidal disruption events (TDEs) by the most massive ($\gtrsim\mr{few}\times 10^7\msun$) black holes (BHs) is suppressed due to direct capture of stars by the event horizon. This suppression effect depends on the shape of the horizon and holds the promise of probing the spin distribution of dormant BHs at the centers of galaxies. By extending the frozen-in approximation commonly used in the Newtonian limit, we propose a general relativistic criterion for the tidal disruption of a star of given interior structure. The rate suppression factor is then calculated for different BH masses, spins, and realistic stellar populations. We find that either a high BH spin ($\gtrsim 0.5$) or a young stellar population ($\lesssim$1 Gyr) allows TDEs to be observed from BHs significantly more massive than $10^8\,\msun$. We call this spin-age degeneracy (SAD). This limits our utility of the TDE rate to constrain the BH spin distribution, unless additional constraints on the age of the stellar population or the mass of the disrupted star can be obtained by modeling the TDE radiation or the stellar spectral energy distribution near the galactic nuclei.
\end{abstract}

\begin{keywords}
Tidal disruption events --- black hole --- general relativity --- transients
\end{keywords}

\section{Introduction}

A prediction of general relativity is that stars can be directly swallowed by the most massive black holes (BHs) without producing an electromagnetic flare \citep{Young1977, Rees1988, Kesden2012, lu17_event_horizon, vanvelzen18_TDE_rate_suppression}. This gives a strong, spin-dependent suppression of observable tidal disruption event (TDE) rate \citep{Kesden2012, Coughlin2022a}. An important goal of the TDE community is to use this suppression effect to constrain the spin distribution of a large number of dormant BHs at the nuclei of galaxies. Given the rapidly growing sample of TDEs enabled by recent surveys \citep{holoien19_TESS_TDE, hung20_AT2018hyz, vanvelzen21_ZTF_TDEs, sazonov21_eROSITA_TDEs, hammerstein22_ZTF_TDEs} and future Vera Rubin Observatory \citep{ivezic19_LSST}, it is very promising to accurately measure the TDE rate as a function of the BH mass, $\Gamma_{\rm TDE}(\mBH)$, provided that the BH masses can be statistically inferred from galaxy scaling relations \citep[e.g., the $\mBH$-$\sigma$ relation, see][]{kormendy13_scaling_relations} that are well calibrated for $\mBH\gtrsim \mbox{few}\times10^7\msun$. 

To approach the goal of constraining the BH spins, in this work we provide an accurate prediction of the observable TDE rate function $\Gamma_{\rm TDE}(\mBH)$, for different BH spins and stellar populations. For a given dimensionless spin and stellar population, this rate function is decomposed into two factors,
\begin{equation}
    \Gamma_{\rm TDE}(\mBH) = f_{\rm TDE}\times \Gamma_{\rm lc},
\end{equation}
where $\Gamma_{\rm lc}$ is the rate at which stars are scattered into the loss cone and $f_{\rm TDE}$ is the fraction of these stars that produce observable electromagnetic flares (and $1-f_{\rm TDE}$ is the fraction of stars that are directly swallowed). Previous studies of the loss-cone dynamics \citep{magorrian99_TDE_rate, wang04_TDE_rate, stone16_TDE_rate, stone20_TDE_rate_review} show that $\Gamma_{\rm lc}$ is likely only a weak function of the BH mass.
Observations also suggest that the TDE rate depend weakly on the BH mass for $\mBH\lesssim 10^7\msun$ for which most loss-cone events produce bright flares \citep[][their Figure 3]{vanvelzen18_TDE_rate_suppression}.

On the other hand, we expect the observable TDE fraction $f_{\rm TDE}$ to drop rapidly at high BH masses as a result of direct captures --- for instance, \citet{Kesden2012} predicted $f_{\rm TDE}(\mBH=10^7\msun)\sim 0.5$ independent of spin and that it drops to $\sim\! 10^{-3}$ (or $\sim \!3\times10^{-2}$) for $\mBH=10^8\msun$ and dimensionless spin parameter $j=0.5$ (or $0.9$). However, \citet{Kesden2012} only considered the case of a solar-like star whereas a realistic stellar population consist of stars of different masses and ages. Moreover, the criterion for tidal disruption in \citet{Kesden2012} is based on the maximum tidal acceleration equaling to the surface gravitational acceleration of the star, but this criterion is not realistic and in fact disagrees with the results of relativistic hydrodynamic simulations by \citet{Ryu2020d} (see Figure \ref{fig:Ryu_et_al_compare} for a comparison).

In the absence of relativistic hydrodynamic simulations for a large number of inclined orbits for spinning BHs, we seek for semi-analytic criteria for tidal disruption that involve the interior structure of the star as well as the relativistic BH spacetime. After briefly introducing the Kerr spacetime in \S \ref{sec:Kerr_basic}, we present our new criteria in \S \ref{sec:TD_criteria}, where we generalize the ``frozen-in'' approximation \citep[as adopted in the Newtonian limit by][]{Lodato2009, stone13_frozen-in, steinberg19_frozen-in} to relativistic geodesics by integrating the tidal acceleration over the orbit, and if a part of the star (in its interior) can be accelerated to the local escape velocity wrt. the stellar center, then we consider the fluid element to be tidally stripped from the star. We find good agreement between this generalized frozen-in approximation and numerical simulations by \citet{Ryu2020d} (which are for Schwarzschild BHs). This motivates us to apply the method to the case of spinning BHs and calculate the observable TDE rate for the highest mass BHs for different stellar populations in \S \ref{sec:TDE_capture_rates}.

It is important to stress upfront that many aspects of the electromagnetic emission from TDEs are poorly understood \citep[the origin of optical emission in particular, see e.g.,][]{piran15_shock_power, metzger16_bright_year, roth16_optical_emission, dai18_unified_model, lu20_optical_emission, bonnerot21_first_light, andalman22_global_GRHD, steinberg22_peak_light}. Currently, there is not a clear mapping between the mass loss from the star to the properties of the emission. In this paper, we adopt a mass-loss fraction of 50\% from the star as a clear-cut boundary between observable and dark TDEs, although our method can be directly applied to other mass-loss fractions. The mass loss fraction is a very strong function of the orbital pericenter radius --- for the cases simulated by \citet{Ryu2020d}, the pericenter radii for mass-loss fraction of 30\% (or 70\%) only different from that for 50\% mass loss by about ten percent, and this would lead to a small change in the observable TDE rate as compared to what is presented in this paper.

The results from our calculations are then presented in \S \ref{sec:result}, including the size of the loss cones for disruption and direct capture for stars of different ages and masses (\S \ref{sec:single_star}). The synthesized TDE rates for the entire stellar population and our proposed spin-age degeneracy
(SAD) are detailed in \S \ref{sec:TDE_rate_KroupaIMF}. We discuss the limitations of our calculations in \S \ref{sec:discussion} and summarize our findings in \S \ref{sec:summary}.

\section{Method}\label{sec:method}

In this section, we describe our criteria for determining the outcome of a star passing by a BH, and how they can be used to calculate the disruption rate and direct capture rate by integrating over the angular momentum and mass distributions of the stellar population.



\subsection{Kerr geodesic and tidal tensor}\label{sec:Kerr_basic}
The spacetime of a rotating BH is given by the Kerr metric, which can be expressed in Boyer-Lindquist coordinates under geometrized units ($G=c=1$) as
\begin{equation}
\begin{split}
        ds^{2} =& -\lrb{1-\frac{2\mBH r}{\Sigma}}dt^{2}+\frac{\Sigma}{\Delta}dr^{2}+\Sigma d\theta^{2} \\
        & +\lrb{r^{2}+a^{2}+\frac{2\mBH ra^{2}}{\Sigma}\sin^{2}\theta}\sin^{2}\theta d\phi^{2} \\
        & -\frac{4\mBH ra\sin^{2}\theta}{\Sigma}dtd\phi,
\end{split}
\label{eq:BL_coor}
\end{equation}
where $\Sigma=r^2+a^2\cos^2\theta,\Delta=r^2-2\mBH r+a^2$, and $\mBH,a$\ are the mass and spin of the BH \citep{Boyer1967}.
In the following we will also frequently use $j=a/\mBH$, which is the dimensionless spin of the BH.

For a main-sequence star passing by a BH, 
its center of mass follows a time-like geodesic.
The radius of the star, which is of the order $\Rsun\approx7\times10^8\,\rm m$, is much smaller than the Schwarzschild radius of the BH with $\mBH\ge10^6\,\msun$:
\begin{equation}
    r_{\rm S}=2\,\rg=\frac{2G\mBH}{c^2}=2.95\times10^9\lrb{\frac{\mBH}{10^6\,\msun}}\,\rm m.
\end{equation}
The geodesic equations are \citep{Carter1968}
\begin{align}
\Delta\dot{t} & =\lrb{r^{2}+a^{2}+\frac{2\mBH ra^{2}}{\Sigma}\sin^{2}\theta}E-\frac{2\mBH ra}{\Sigma}L_{z},\\
\Delta\dot{\phi} & =\lrb{1-\frac{2\mBH r}{\Sigma}}\frac{L_{z}}{\sin^{2}\theta}+\frac{2\mBH ra}{\Sigma}E,\\
\Sigma^{2}\dot{\theta}^{2} & =Q+\cos^{2}\theta\lrsb{(E^{2}-1)a^{2}-\frac{1}{\sin^{2}\theta}L_{z}^{2}},\\
\Sigma^{2}\dot{r}^{2} & =\lrsb{E(r^{2}+a^{2})-aL_{z}}^{2}-\Delta\lrsb{Q+(L_{z}-aE)^{2}+r^{2}},
\end{align}
where $E$ is the specific energy, $L_z$ is the specific angular momentum along the black hole spin axis, and $Q$ is the Carter constant.
Note that $E,L_z,Q$ are all constants of motion and are conserved along the geodesic.
Far from the BH ($r\gg \rg$), the Carter constant is related to the total specific angular momentum $L$ by
\begin{equation}
    Q=L^2-L_z^2.
\end{equation}
The above geodesic equations are numerically solved using the code developed by \citet{Rauch1994}.
Furthermore, we are only interested in the stellar trajectories that will come close to the BH and potentially produce TDE.
As the initial kinetic energy of the star is negligible compared to the work done by tidal forces in the BH's frame as the star reaches near the pericenter, we simply set $E=1$ in all the calculations.

The tidal disruption can be viewed as the consequence of the differential motion of fluid elements in the star induced by the tidal forces of the BH.
Due to the smallness of the star compared to the BH, it is convenient to describe the process in the local frame of the star, using the tidal tensor.
In the local inertial frame of the center of a free-falling star (Fermi Normal Coordinates), the motions of fluid elements will follow the equation of geodesic deviation in the absence of other forces:
\begin{equation}\label{eq:geodesic-deviation}
    \frac{d^{2}\chi^{i}}{d\tau^{2}}=-C_{ij}\chi^{j},
\end{equation}
where $\tau$ is the proper time of the geodesic, $\chi$ is the displacement of the fluid element from the stellar center, and $C_{ij}$ is the tidal tensor.
The tidal tensor is a $3\times3$ symmetric matrix described in Appendix \ref{sec:tidal_tensor} and see \cite{Marck1983} for more details.

\subsection{Criteria for tidal disruption}\label{sec:TD_criteria}
Since we are interested in not the details of each TDE but the overall rate, we adopt a critical mass loss fraction of 50\% as the threshold for luminous TDEs. In the following the star is said to be disrupted only if it loses more than 50\% of its mass during the pericenter passage. Our rate calculation is not sensitive to this choice, since the mass loss fraction is a very steep function of the pericenter radius of the stellar trajectory \citep{guillochon13_TDE_hydro, law-smith20_TDE_hydro, Ryu2020a}. In fact, our method can be directly applied to other choices as well (e.g., 30\% mass loss).

Whether the star can be disrupted depends on both the strength and the working time of the BH's tidal forces.
Without full-scale hydrodynamic simulations, we attempt to devise the criteria of tidal disruption that capture these two aspects of tidal forces.
Our criteria have two parts.
First, motivated by the frozen-in approximation, we calculate the work done by tidal forces by integrating the geodesic deviation equation to obtain ``maximum differential velocity'' $\Delta v_{\rm max}$.
As explained below, the value of $\Delta v_{\rm max}$ for a given stellar geodesic depends on the initial radius $r_0$ from which we start our integration. Second, the star must fill up its Roche Lobe (described in Fermi Normal Coordinates) at radii $r < r_0$ in order for it to lose mass.

In the following, we elaborate on the concept of relativistic Roche lobe and how the maximum differential velocity $\Delta v_{\rm max}$ is calculated, and based on these, we then construct the criteria for tidal disruption. A comparison of our criteria to the numerical simulation results by \cite{Ryu2020a} is then provided as a test of the validity for the Schwarzschild case. Since our tidal disruption criteria are based on the time-dependent tidal tensor in the Fermi Normal frame, the formalism can be directly applied to any geodesics in the Kerr metric. We stress that our criteria are only approximate (certainly not perfect) and the validity needs to be strictly tested against future hydrodynamic simulations in the Kerr spacetime. However, given the large computational cost of such numerical simulations, our method provides the best-effort TDE rate predictions on the high BH-mass end before such extensive hydrodynamic simulations are carried out.

\subsubsection{Maximum differential velocity}\label{sec:max_deltav}
The frozen-in approximation \citep{stone13_frozen-in} assumes that a star is unperturbed before reaching the Newtonian tidal disruption radius $r_{\rm T} =R_*(M_{\rm BH}/M_*)^{1/3}$ ($R_*$ and $M_*$ being the stellar radius and mass), at which the star is rapidly torn apart and then the fluid elements start to free-fall according to the gravitational field of the BH.

\begin{figure}
    \centering
    \includegraphics[width=\linewidth]{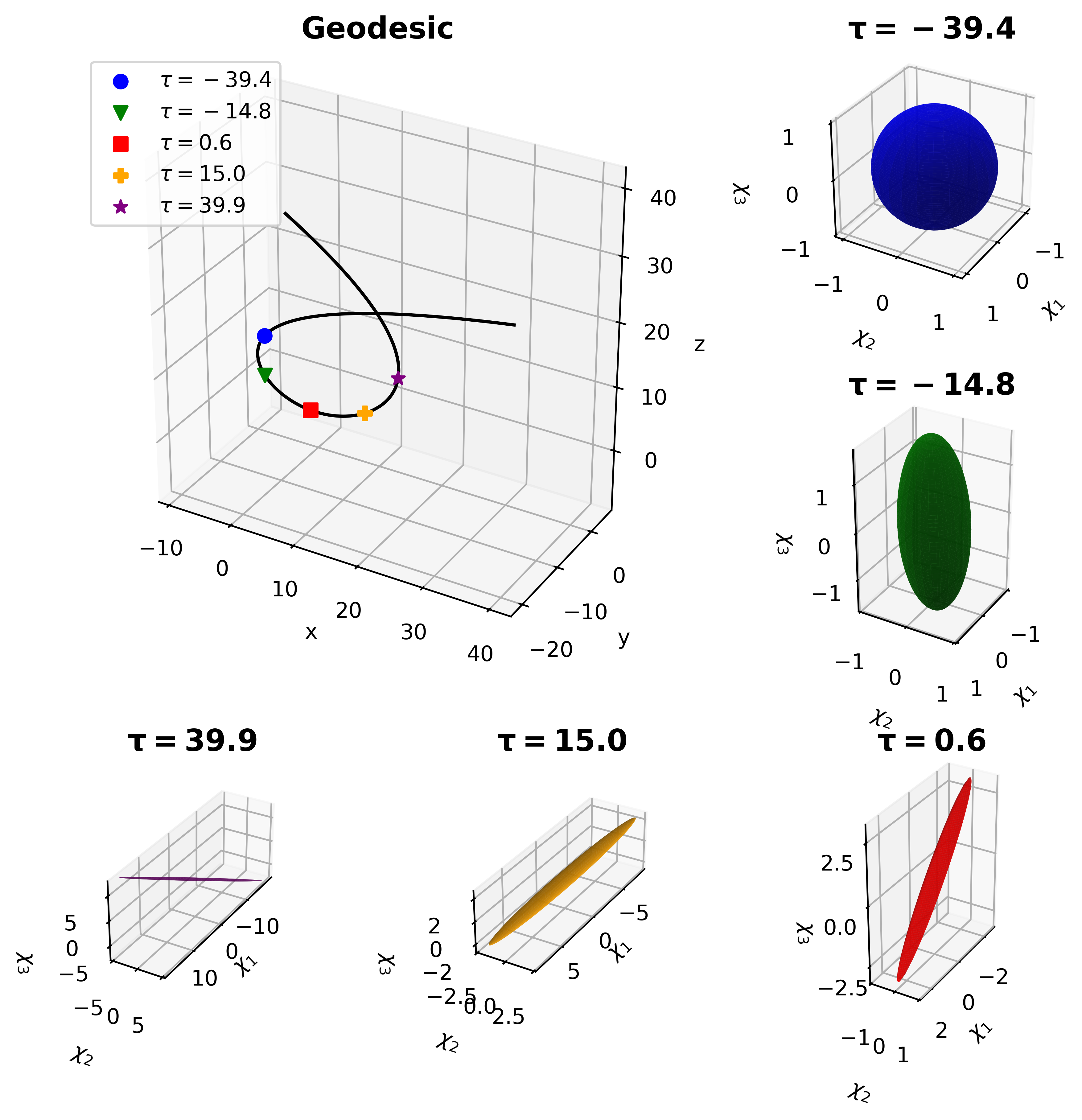}
    \caption{
    An example of the geodesic and half-mass surface distorted by tidal forces under the frozen-in approximation.
    The top-left panel shows a parabolic geodesic in the Kerr metric of $j=0.9$. In Boyer-Lindquist coordinates and natural units, the geodesic starts at $r=1000,\theta=\pi/2,\phi=0$, with $L=4.62$ and $L_z/L=-0.5$.
    Five markers are placed on the geodesic to represent the different stellar proper times $\tau$ ($\tau=0$ at the pericenter).
    Blue circle marker is the point the star center enters $r_0=13$; purple star marker is the point the star center leaves $r_0=13$; red square marker is a point close to the pericenter.
    We assume that, as soon as the star enters $r_0=13$, the fluid elements at the half-mass radius start to free-fall under the influence of tidal forces (eq. \ref{eq:geodesic-deviation}), causing the sphere to be distorted.
    Such distortions are shown in the remaining panels using Fermi Normal Coordinates.
    The length unit in Fermi Normal Coordinates is normalized to the original half-mass radius $\Rhm$.
    Note that the surfaces shown in this figure are shown using a very high resolution sampling grid across the half-mass surface (and triangulation), whereas in our production-run
    calculations of $\Delta v(r_0)$ (see Section \ref{sec:max_deltav}), a much coarser (yet sufficient) grid is used to reduce the computational cost.
    }
    \label{fig:geodesic_surface}
\end{figure}

In our relativistic treatment of tidal disruption, we assume the star to be unperturbed before reaching radius $r_0$, which is not equal to the Newtonian tidal disruption radius $r_{\rm T}$.
At $r=r_0$, the self-gravity and pressure forces of the star are overwhelmed by the tidal forces and fluid elements in the star begin to free-fall. We track the free-fall of the fluid elements that originate at the half-mass radius $\Rhm$ of the star by solving the equation of geodesic deviation explicitly. The choice of the half-mass radius is based on our use of 50\% mass loss as the threshold of tidal disruption, and the exact value of $\Rhm$, which depends on both the age and the mass of the star, is calculated from a stellar model obtained by Modules for Experiments in Stellar Astrophysics \citep[MESA,][]{Paxton2011, Paxton2013, Paxton2015, Paxton2018, Paxton2019}.
An illustration of the free-fall of the fluid elements is provided in Figure \ref{fig:geodesic_surface}, where we show an inclined geodesic in the Kerr metric and how the sphere of the half-mass radius get distorted with time under the influence of tidal forces described by eq. (\ref{eq:geodesic-deviation}).

These fluid elements are uniformly sampled on a sphere of radius $\Rhm$ according to their angular positions $(\theta, \phi)$ in the local frame of the star. The number of sampling grid points is $5\times5$ and the grid is uniform in the range of $\cos\theta\in(-1, 1),\phi\in(0,2\pi)$, which we show to be sufficient based on our convergence test. In order for the star to lose 50\% of its mass, these fluid elements must be able to produce a sufficiently large differential velocity $\Delta v$ so as to break free from the self-gravity of the star. Based on each of the fluid elements initially at $(\theta, \phi)$ at radius $\Rhm$ from the stellar center, we calculate the largest differential velocity by the time the star exits from the sphere of radius $r_0$ from the BH and refer to it as $\Delta v(r_0) \equiv \max \lrsb{\Delta v(\theta, \phi)}$.

The choice of the initial radius coordinate $r_0$ is an issue in this approach. However, we find that the following method produces results that are in good agreement with those from the hydrodynamic simulations by \citet{Ryu2020a}. Step (1): we evaluate $\Delta v(r_0)$ on a dense grid\footnotemark of $r_0$ and then choose the critical $r_0$ that maximizes the function of $\Delta v(r_0)$. The maximum differential velocity $\Delta v_{\rm max}$ obtained in Step (1) characterizes the maximum work that can be done by the BH's tidal forces on a given fluid element. Step (2): we compute the Roche Lobe volume $V_{\rm RL}(r_0)$ as a function of $r_0$ along the geodesic and restrict the choice of $r_0$ in Step (1) by the requirement of $V_{\rm RL}(r_0) < V_*$ (where $V_* = 4\pi R_*^3/3$ is the volume of the unperturbed star). This is a conservative restriction, which means that the star will lose mass at radii $r < r_0$.

\footnotetext{The grid is composed of 20 grid points and equally spaced in the variable $\log_{10}r_0^\prime\in[-3, 0.3]$, where $r_0^\prime=(r_0-\rp)/\rp$.}

\begin{figure}
    \centering
    \includegraphics[width=\linewidth]{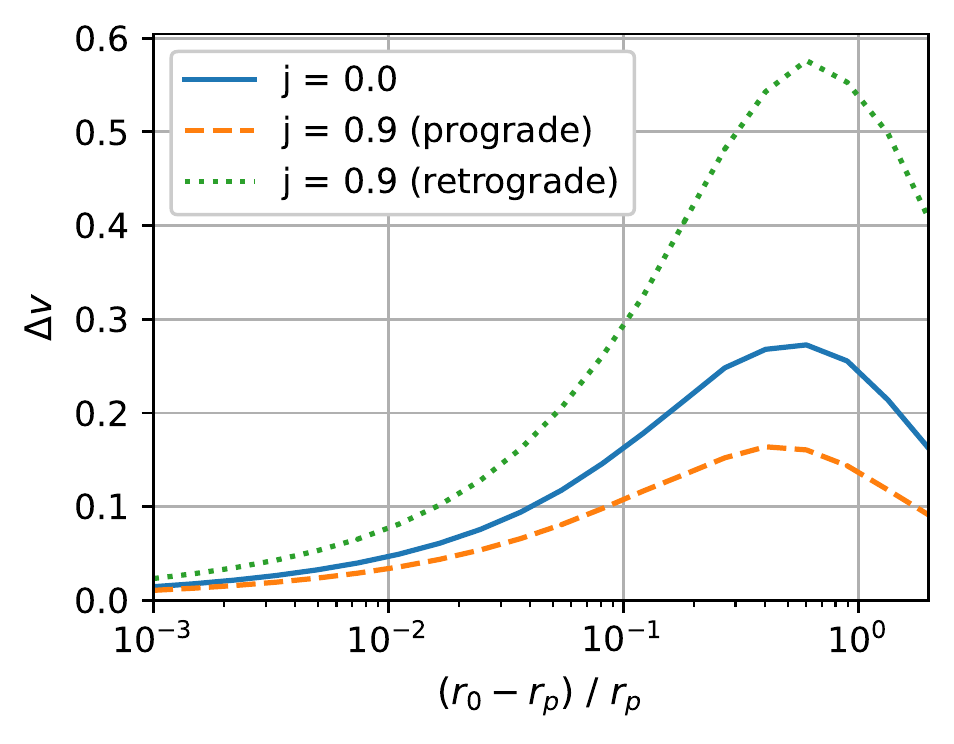}
    \caption{
    Maximum differential velocity $\Delta v$ as the function of the tidal radius $r_0$ for three geodesic in the Kerr metric of different spins $j$.
    All geodesics have $L=4.62$, which for the Schwarzschild metric (blue solid line) corresponds to $\rp=8.0\,\rg$.
    For the geodesic of prograde (retrograde) orbit plotted in orange dashed (green dotted) line, it has $l_z=L_z/L=0.5\ (-0.5)$ and $\rp=8.7\ (7.1)\,\rg$.
    The value of $\Delta v$ is calculated from an initial radius of $\abs{\boldsymbol{\chi}_0}=1$ (normalized such that $\Rhm=1$), in the natural units $G=\mBH=c=1$.
    }
    \label{fig:deltav_r0}
\end{figure}

In Figure \ref{fig:deltav_r0}, we show $\Delta v(r_0)$ for parabolic, inclined geodesics of $L=4.62$ and $|L_z/L| = 0.5$ (in natural units) in the Kerr metric for different BH spins $j=0$ and $0.9$. The differential velocity $\Delta v$ depends on the initial radius $r_0$ in the following way. For large enough $r_0$, $\Delta v$ decreases with $r_0$ because of the declining strength of the BH's tidal forces; whereas as $r_0$ approaches $\rp$, $\Delta v$ vanishes rapidly because of the decreased time for the tidal forces to do work. For the example of $L=4.62$ in the Schwarzschild metric ($j=0$), we find that differential velocity reaches its maximum $\Delta v_{\rm max}$ at $r_0/\rp\approx1.7\rg$.
Figure \ref{fig:deltav_r0} also shows the dependence of the work done by tidal forces on the BH spin. For fixed $L$, the prograde (retrograde) orbit in the spinning BH will experience smaller (larger) tidal forces and has a lower (higher) value of $\Delta v$.
We will return to this point in \S \ref{sec:single_star}.


In the next subsection, we discuss the requirement that the star must fill up its Roche lobe at radii $r < r_0$ in order for it to lose mass.



\subsubsection{Relativistic Roche Lobe}

To complete our discussion on maximum differential velocity $\Delta v_{\rm max}$, we provide a constraint on the initial radius $r_0$. In the face of the enormous challenge of modeling the hydrodynamic effects including the stellar interior structure and the tidal acceleration history along the geodesic \citep[see][for a review]{Rossi2021}, we only seek to place a limit on the potential range of $r_0$ using the concept of the Roche lobe.

In the studies of binary systems, the Roche lobe is the largest volume that a star can occupy without losing mass to its companion \citep{Paczynski1971}.
A similar concept can be applied to the mass loss of the star under the effect of tidal forces.
For a tidal tensor $C_{ij}$, the motion of a fluid element in the Fermi Normal frame is governed by the tidal potential $\Phi_{\rm tide}$, the star's original gravitational potential, as well as pressure forces. Hydrostatic equilibrium is reached when the isobars are coincident with the equipotential surfaces (including both tidal and self-gravity potentials). There is a critical value of potential $\Phi_{\rm crit}$, above which the equipotential surfaces are no longer closed around the star. The region enclosed within the equipotential surface corresponding to $\Phi_{\rm crit}$ is the largest volume that a star can have before starting to lose mass, under the assumption of hydrostatic equilibrium.

We calculate the volume of the Roche Lobe $\VRL(r)$ along the geodesic at different radii. The detailed (but straightforward) calculations of the shape of the Roche lobe and its volume $\VRL$ are provided in Appendix \ref{sec:Roche_lobe}. It should be noted that here we assume that the star is rotating at angular frequencies that are much below the Keplerian frequency at the stellar surface $\sqrt{GM_*/R_*^3}$, and under this assumption, we ignore the centrifugal forces in the Roche potential. This is appropriate since most stars that are scattered into the loss cone are expected to be slow rotators. Since a necessary condition for tidal disruption is mass loss from the star, the volume of the Roche lobe sets a constraint on the value of $r_0$ by requiring
\begin{equation}
    \VRL(r_0) < V_* = 4\pi R_*^3/3,
\end{equation}
where $V_*$ is the volume of the unperturbed star. This provides a maximum value of $r_0$, which is referred to as $r_{\rm 0,max}$ hereafter. Beyond this radius, the BH's tidal forces are too weak to induce mass loss from the star and the frozen-in approximation is unlikely to apply. There are non-plunging geodesics where $r_{\rm 0,max}$ does not exist, because $\VRL(r)$ is greater than $V_*$ everywhere along the geodesic. In those cases, the star is not tidally disrupted for that geodesic.



\subsubsection{Combination of maximum differential velocity and Roche lobe}
Equipped with the differential velocity as a function of initial radius $\Delta v(r_0)$ (already maximized over all fluid elements at $R_{\rm hm}$ from the stellar center) as well as the maximum initial radius $r_{0,\max}$, we can now fully describe our criteria for tidal disruption. We evaluate $\Delta v(r_0)$ on a dense grid of $r_0\in [\rp, r_{\rm 0,max}]$. The maximum value of $\Delta v(r_0)$ within the range $r_0\in[\rp,r_{\rm 0,max}]$ is the largest differential velocity that the fluid elements originated from $\Rhm$ can achieve and is denoted as $\Delta v_{\max}$.
The star is then only classified as disrupted if $\Delta v_{\max}$ is greater than the critical velocity $v_{\rm vir}$, where
\begin{equation}\label{eq:virial_velocity}
    v_{\rm vir} = \sqrt{G\mstar/2\over \Rhm}.
\end{equation}
which is the virial velocity for a fluid element at the radius $\Rhm$.

\begin{figure}
    \centering
    \includegraphics[width=\linewidth]{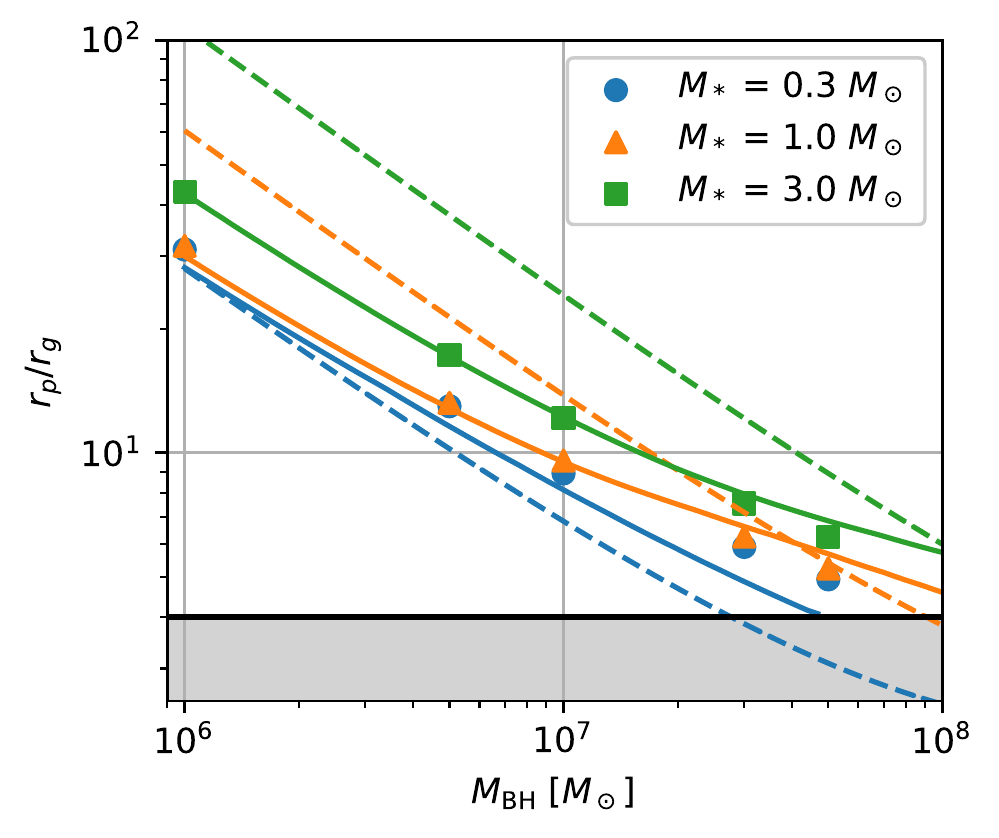}
    \caption{
    The maximum pericenter that is able to produce the tidal disruption with 50\% mass loss for different SMBH and stellar masses.
    The SMBHs are non-rotating ($j=0$) and the spacetime is simplified into Schwarzschild metric.
    The markers are the data points interpolated from Figure 6 of \protect\cite{Ryu2020d}.
    The solid lines are the prediction from our criteria and the dashed lines are the prescription of the tidal disruption used in \protect\cite{Kesden2012}.
    To ensure a fair comparison, the stellar properties are computed from MESA
    and taken at the half age of the main sequence.
    The horizontal black line indicates the minimum pericenter $4\,\rg$ that a star can reach without being captured by the SMBH.
    The gray-shaded region indicates the pericenters inaccessible by the stars.
    }
    \label{fig:Ryu_et_al_compare}
\end{figure}


We test the validity of our criteria of tidal disruption by computing the maximum pericenter radius that can produce tidal disruption with 50\% mass loss for non-spinning BHs. Our results are compared with those from the hydrodynamic simulations by \cite{Ryu2020a,Ryu2020d} and shown in Figure \ref{fig:Ryu_et_al_compare}. We interpolate the data points in Figure 6 of \cite{Ryu2020d} to get the pericenter of the stellar trajectory that can produce the tidal disruption with 50\% mass loss.
For all three stellar masses, $0.3,1.0,3.0\,\msun$, our criteria match the result of the hydrodynamic simulations at low BH masses ($\sim\,10^6\,\msun$).
The value of maximum pericenter from our criteria is slightly higher for $\mstar=3.0\,\msun$ and lower for $\mstar=0.3\,\msun$ than the result of hydrodynamic simulation at high BH masses ($\sim\,10^{7.5}\,\msun$). Such discrepancies are expected from the approximate nature of our treatment.

In the earlier study of tidal disruption by Kerr BHs, \cite{Kesden2012} proposed a tidal disruption criterion based on the comparison between the strength of the maximum tidal force (given by the maximum eigenvalue of the tidal tensor) and the surface gravity of the star at the pericenter.
From Figure \ref{fig:Ryu_et_al_compare}, we see that the critical pericenter radii from \cite{Kesden2012}'s prescription has a nearly power-law dependence on the BH mass, $\rp\propto M^{-1/3}$ (close to that expected from the Newtonian prescription), which fails to reproduce the hydrodynamic results at high BH masses. In particular, the hydrodynamic results show a flattening in the critical pericenter radii at high BH masses, and this flattening is due to the longer working time of the tidal forces in those cases, which are not captured by \cite{Kesden2012}'s criterion based on the comparison of instantaneous forces. Moreover, \cite{Kesden2012}'s prescription does not make use of the information on the stellar interior structure. 
Stars slightly heavier than about $1\,\msun$ are more difficult to be tidally disrupted than lower mass stars as the interior structure transits from convective to radiative envelope without much increase in the stellar size.
Ignoring the interior structure leads to incorrect critical pericenter radii even for low-mass BHs.

Therefore, despite some small discrepancies, we regard our criteria as a significant improvement from \cite{Kesden2012} as we use a more sophisticated, physically motivated treatment on relativistic effects and the stellar structure.

\subsection{Rates of direct captures and observable tidal disruption events}\label{sec:TDE_capture_rates}


With the criteria for tidal disruption in hand, we then proceed to calculate the rates of direct captures and tidal disruptions based on the orbital angular momentum and mass distributions of stars that are scattered into the loss cone.

All the stars coming close to the BH pass through the surface of a sphere of radius $\rinit\ggg \rg$ centered at the BH. Following \citet{Kesden2012}, we assume the ``full loss-cone'' regime in this work \citep[as also adopted by][]{Coughlin2022a}, meaning that the velocity distribution on the surface of the $\rinit$ sphere is taken to be uniform. Our formalism can be directly applied to any angular momentum distribution for the stars scattered into the loss cone. We leave a detailed exploration of other angular momentum distributions \citep[e.g., allowing a certain fraction of galaxies to be in the ``empty loss-cone'' regime, see][]{stone16_TDE_rate} to a future work.

Let $n$ and $v$ stand for the number density and velocity of stars on the surface of the sphere at radius $\rinit$.
The differential rate of the stars entering the sphere in terms of specific angular momentum is given by
\begin{equation}\label{eq:diff_rate}
    \frac{\partial^2\Gamma}{\partial L\partial l_z}=\frac{\pi nL}{v},
\end{equation}
where $L$ is the total specific angular momentum of the star and $l_z=L_z/L\in[-1,1]$ is the fractional angular momentum projected in the direction of the BH spin axis.
The derivation of the above expression is provided in Appendix \ref{sec:differential_rate}.
Note that to fully specify a geodesic, the initial radius $\rinit$ and polar angle $\thetainit$ (wrt. the BH spin axis) are needed in addition to $L$ and $l_z$. Since tidal interactions are only important near the BH, the results do not depend sensitively on the choice of $\rinit$, as long as it is sufficiently large. Based on this, we fix $\rinit=1000\,\rg$. We also find that, for $\rinit\gg\rg$, the strength of the BH's tidal forces at small radii are rather insensitive of the initial polar angle $\thetainit$, with maximum fractional difference in the eigenvalues of the tidal tensor typically less than 10\% (only in the most extreme rare cases of $j=0.99$ and nearly plunging geodesics, the maximum difference reaches to about 20\%).
To save computational cost, we fix $\thetainit=\pi/2$ for all geodesics. The fact that the tidal forces at small radii are insensitive to $\thetainit$ also allows us to integrate the differential rate over $\thetainit$ to obtain eq. (\ref{eq:diff_rate}).

\begin{figure*}
    \centering
    \includegraphics[width=\textwidth]{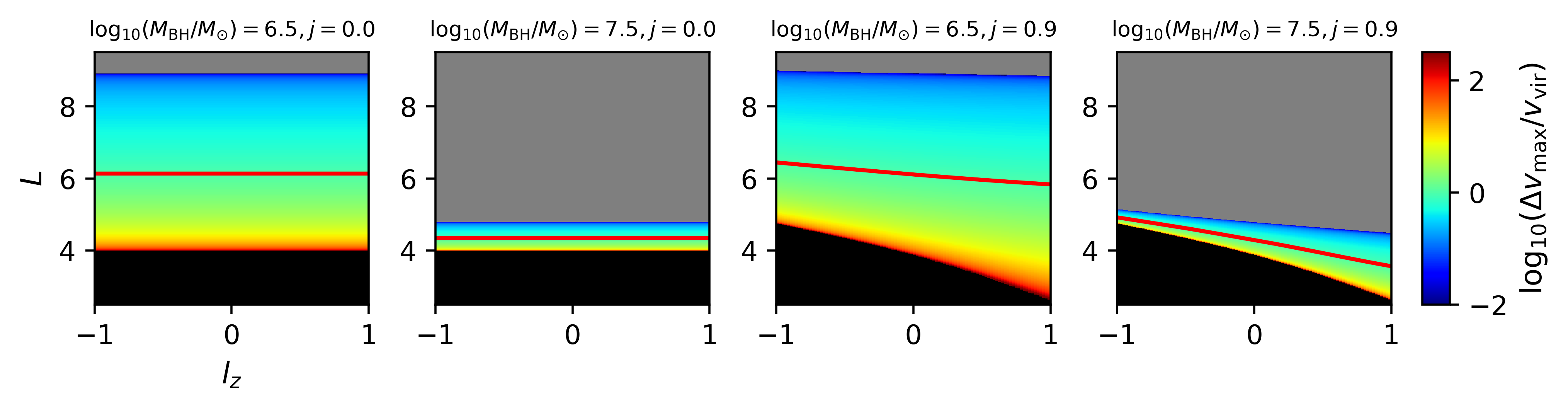}
    \caption{
    The ratio $\Delta v_{\max}/v_{\rm vir}$ (see Section \ref{sec:TD_criteria} for their definitions) in the $L,l_z$ parameter space for different BH masses $\log_{10}(\mBH/\msun)=6.5,7.5$ and spins $j=0.0,0.9$.
    The calculations are for a $1.0\,\msun$ main-sequence star of age $1.0\,\Gyr$.
    The angular momentum $L$ is expressed in natural units $G=\mBH=c=1$.
    The red line in each panel indicates the location of $\LTD$, for which $\Delta v_{\max}=v_{\rm vir}$.
    The black region at the bottom indicates the geodesics directly captured by the BH.
    Between the red line and black region, the star can produce an observable TDE.
    The gray region indicates where the star does not fill up its Roche lobe and hence there will be no mass loss.
    }
    \label{fig:delta_v}
\end{figure*}

For a given $l_z$ (specifying the inclination angle of the orbit), there are two important values of the total angular momentum, $\Lcapt(l_z)$ and $\LTD(l_z)$, which are the critical values of the total angular momentum for direct capture and tidal disruption, respectively. The value of $\Lcapt$ only depends on the constants of motion and can be calculated numerically \citep[see also][]{Coughlin2022a}. On the other hand, $\LTD$ also depends on the stellar properties and only exists if the BH's tidal forces are strong enough to produce tidal disruption.
We further define the loss cone angular momentum $\Llc$ as the maximum of $\Lcapt$ and $\LTD$, if the latter exists, i.e.
\begin{equation*}
    \Llc =\max\lrb{\Lcapt, \LTD}.
\end{equation*}
With the above definitions, the TDE rate $\GammaTDE$ of a stellar population of fixed stellar mass and age is given by
\begin{align}
    \GammaTDE = & \int_{-1}^1\int_{\Lcapt}^{\Llc}\frac{\pi nL}{v}\,dLdl_z\nonumber\\
               = & \frac{\pi n}{2v}\int_{-1}^1 \lrsb{\Llc^2(l_z)-\Lcapt^2(l_z)}\,dl_z.
\end{align}
Similarly, the capture rate can be expressed as
\begin{align}
    \Gammacapt = & \int_{-1}^1\int_0^{\Lcapt}\frac{\pi nL}{v}\,dLdl_z\nonumber\\
               = & \frac{\pi n}{2v}\int_{-1}^1 \Lcapt^2(l_z)\,dl_z.
\end{align}
The sum of $\GammaTDE$ and $\Gammacapt$ is the rate at which the stars enter the loss cone
\begin{equation}
    \Gammalc = \GammaTDE + \Gammacapt.
\end{equation}
To calculate $\Gammalc,\GammaTDE,\Gammacapt$ and thus $\LTD$, we sample the geodesic on the grid of $L,l_z$. 
For $j=0.0$, the grid number is 8000 in the range of $L\in[4+10^{-5},14.5]$ and uniform in log scale.
For $j\ne 0.0$, the grid number is $1000\times100$ in the range of $L\in[\Lcapt(l_z=1.0)+10^{-5},14.5]$ and $l_z\in(-1,1)$. The grid is uniform in log scale of $L$ and in linear scale of $l_z$.

The above ``monochromatic'' (for single $\mstar$ and $\tage$) TDE and capture rates depend on the mass $\mBH$ and spin $j$ of the BH, as well as the mass $\mstar$ and age $\tage$ of the star.
The observed TDE rates from the galactic nuclei are then obtained by averaging $\GammaTDE(\mBH,j,\mstar,\tage)$ over a given stellar population of different $\tage,\mstar$.

The stellar populations in extragalactic nuclei, especially the stars within a few parsecs from the BH, are uncertain due to the lack of observational constraints. In the following, we adopt a single stellar population of a given age following the \citet{Kroupa2001} initial mass function (IMF). Realistically, the stars that are scattered into the loss cones in a given galactic nucleus were formed at different epochs throughout the history of the galaxy, and different galaxies may have different stellar populations near the centers depending on their evolutionary history. Our results for a single stellar population of different ages can be statistically combined to mimic any realistic stellar populations.

We use MESA to model the evolution of solar-metallicity stars on a sufficiently wide mass grid\footnote{The stellar mass $M_*$ is sampled uniformly in the log scale in the range of $M_*\in[10^{-1},10^{0.71}]$ with 13 grid points.} and record the structures of all stars below the main-sequence turn-over mass $M_{\rm max}(t_{\rm age})$ at a given age $t_{\rm age}$.
The stellar population at this age can be approximately described as a truncated Kroupa IMF within the mass range $M_{\min}\le\mstar\le M_{\max}$, where $M_{\min}=0.085\,\msun$ is the minimum stellar mass\footnote{TDEs by stars with even lower masses, mostly brown dwarfs, are fainter (due to smaller energy budget) and faster fading (due to shorter fallback timescale), so they are not expected to dominate the observed rate in current surveys. } considered in this work. It is further assumed that the stellar populations of different masses have the same initial velocity $v$ and only differ in their number densities $n$. With these assumptions, the stellar population-averaged TDE rate and the direct-capture rate are given by
\begin{align}
    \lara{\Gamma_{k}} =  \frac{\int_{M_{\min}}^{M_{\max}(\tage)}\Gamma_{k}(\mBH,j,\mstar,\tage)\frac{dN}{d\mstar}\,d\mstar}{\int_{M_{\min}}^{M_{\max}(\tage)}\frac{dN}{d\mstar}\,d\mstar},
\end{align}
where the subscript $k=\mr{TDE}$ (for observable TDEs) or $\rm{capt}$ (for direct captures), and $dN/d\mstar$ is the properly normalized Kroupa IMF. Note that the averaged TDE and capture rates, $\lara{\GammaTDE}$ and $\lara{\Gammacapt}$, are both functions of $\mBH,\ j,\ \tage$. The total population-averaged rate of loss-cone scatterings is
\begin{equation}
    \lara{\Gammalc} = \lara{\GammaTDE} + \lara{\Gammacapt}.
\end{equation}

We further define the observable TDE fraction $\fTDE$ as the fraction of stars that are scattered into the loss cone
\begin{equation}\label{eq:fTDE}
    \fTDE=\frac{\GammaTDE}{\Gammalc},
\end{equation}
which does not depend on the stellar number density $n$ and velocity $v$ as they are canceled in the expression. Our goal is to calculate the observable TDE fraction $\fTDE$ as a function of $\mBH, \ j, M_*, \tage$ (i.e., for different BH and stellar properties). Finally, the stellar population-averaged observable TDE fraction $\lara{\fTDE}$ can similarly be defined based on the averaged rates
\begin{equation}\label{eq:fTDE_avg}
    \lara{\fTDE}=\frac{\lara{\GammaTDE}}{\lara{\Gammalc}},
\end{equation}
which can be directly compared with observations provided that we know the total loss-cone scattering rate $\lara{\Gammalc}$ (which depends weakly on BH masses, see below).



\section{Results}\label{sec:result}


In this section we present the TDE rate calculation based on the method in Section \ref{sec:method}.
We first calculate the TDE rate fraction $\fTDE$ of a single star of given initial mass and age in \S \ref{sec:single_star}, and then go on to show the averaged TDE rate fraction $\lara{\fTDE}$ for a stellar population of a given age in \S \ref{sec:TDE_rate_KroupaIMF}.

\subsection{Test case of a single star}\label{sec:single_star}

In this subsection, we consider the TDE fraction $\fTDE$ as a function of BH mass and spin for a fixed $1\,\msun,1\,\Gyr$-old main-sequence star.

\subsubsection{TDE and direct-capture cross-sections}

\begin{figure}
    \centering
    \includegraphics[width=0.9\linewidth]{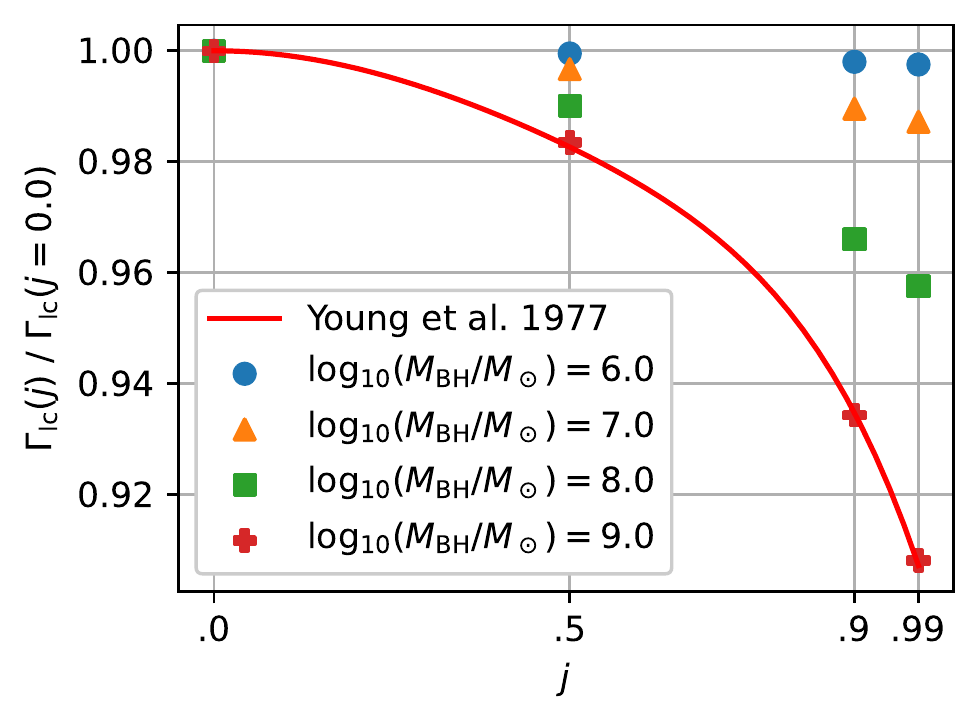}
    \caption{
    The spin-dependent loss-cone scattering rate $\Gammalc(j)$ normalized to the value for a non-spinning BH $\Gammalc(j=0.0)$, for a $1.0\,\msun,1.0\,\Gyr$ main-sequence star. The red curve is the analytical fit given by eq. B2 in \citet{Young1977}.
    }
    \label{fig:Gammalc_j_Gamma_lc_0}
\end{figure}

There are generally three distinct regions in the parameter space $L,l_z$ for the geodesics of the star.
(1) For sufficiently large $L$, the star stays far from the BH and is neither tidally disrupted nor captured; 
(2) For sufficiently small $L$, the star directly plunges into the BH and there is no observable TDE;
(3) In between these two regions, the star is tidally disrupted by the tidal force without being captured and produces an observable TDE.
In the following, we use the criteria developed in Section \ref{sec:TD_criteria}, which is based on the ratio between the maximum differential velocity $\Delta v_{\max}$ (given by the work done by tidal forces) and the virial velocity $v_{\rm vir}$ (eq. \ref{eq:virial_velocity}) at the half-mass radius of the star, to examine how these regions are influenced by BH mass $\mBH$ and spin $j$.

\begin{figure*}
    \centering
    \includegraphics[width=0.9\textwidth]{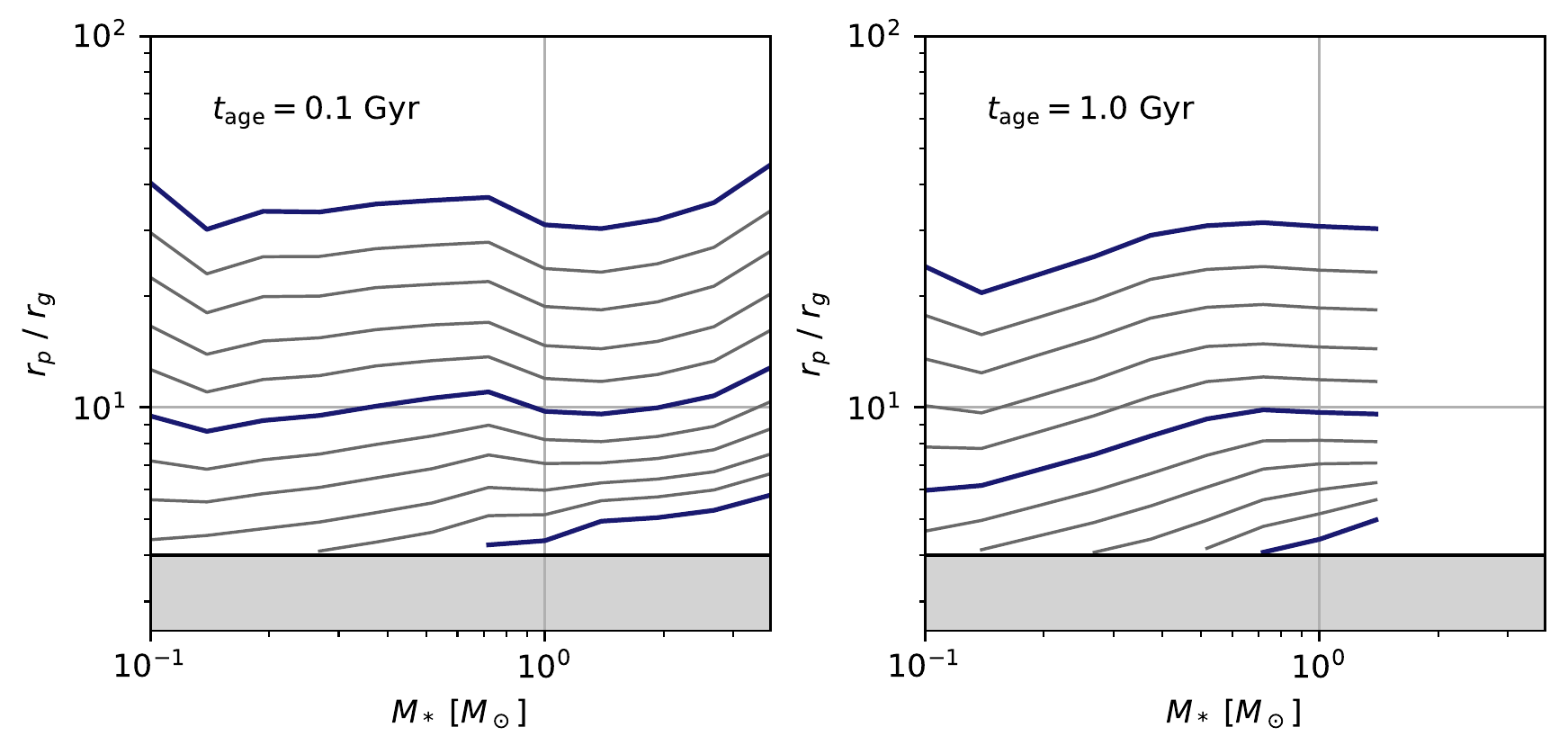}
    \caption{
    The maximum pericenter radii for TDEs as a function of stellar mass and BH mass, for two different stellar ages $\tage=0.1\rm\, Gyr$ (left panel) and $1\rm\, Gyr$ (right panel). The BH is non-spinning ($j=0$).
    In each panel, different curves, from top to bottom, represent the increasing BH masses, starting from $\mBH=10^6\,\msun$ with a step size $\Delta \log_{10}(\mBH/\msun)=0.2$. The cases of $\mBH=10^6,10^7,10^8\,\msun$ are highlighted with dark blue lines.
    The horizon black line at $\rp=4\,\rg$ indicates the minimum pericenter below which the star would be captured by the BH.
    }
    \label{fig:rTD_mstar}
\end{figure*}

Figure \ref{fig:delta_v} shows the ratio $\Delta v_{\max}/v_{\rm vir}$ in the parameter space of $L,l_z$ for two different BH masses $\log_{10}(\mBH/\msun)=6.5,7.5$ and two different spins $j=0.0,0.9$.
In each panel, the capture region in the parameter space $L,l_z$ is colored in black.
For each inclination angle as specified by $l_z$, the upper bound of the capture region is $\Lcapt$, which independent of the stellar properties, and when expressed in the natural units, independent of BH mass $\mBH$.

For $L > L_{\rm capt}(l_z)$, we calculate the maximum differential velocity $\Delta v_{\max}$ based on the method described in Section \ref{sec:TD_criteria}. The value of $\Delta v_{\max}$ contains information on the stellar structure and the relativistic tidal forces.
The star is only considered to be tidally disrupted when $\Delta v_{\max}$ exceeds the virial velocity $v_{\rm vir}$ at the star's half-mass radius.
The critical angular momentum $\LTD$ for which $\Delta v_{\max}=v_{\rm vir}$ is marked by a red solid line.
Between $\LTD$ and $\Lcapt$, the star experiences sufficiently strong tidal forces that lead to an observable TDE. For very large angular momenta $L$, tidal forces are so weak that mass loss from the star is not possible and this region is shaded in gray.

The presence of the BH spin leads to asymmetry in prograde ($l_z>0$) and retrograde orbits ($l_z<0$).
For a fixed total angular momentum $L$, the retrograde orbits experience stronger tidal forces and have larger values of $\Delta v_{\max}$ than the prograde orbits. This effect is reflected in the value of $\LTD$, which is larger for retrograde orbits and smaller for prograde orbits.
The asymmetry of $\LTD(l_z)$ between prograde and retrograde orbits are more important for large BH masses, where $\LTD$ can be very close to $\Lcapt$, meaning that TDEs can only occur close to the capture region.
The strongest tidal forces experienced by a given star, however, occur not in retrograde orbits but in prograde orbits.
This is because the BH spin lowers the value of $\Lcapt$ for prograde orbits, and the star can reach closer to the BH without being captured.

\subsubsection{Weak spin-dependence of the loss-cone cross-section}

For a given inclination $l_z$, the BH spin changes the critical angular momenta for capture, $\Lcapt$, and for TDE, $\LTD$. In this subsection, we show that when averaged over all inclination angles (assuming that stars at large distances are not aware of the BH's spin direction), the total loss-cone rate $\Gammalc$, the rate at which stars are scattered into the loss cone, is nearly independent of the BH's spin for all BH masses relevant for observable TDEs ($\mBH< 10^9\msun$). 

Such a weak dependence is expected for low BH masses for which $\LTD\gg \Lcapt$ or $\Gammacapt \ll \GammaTDE$, because most TDEs occur far from the BH's horizon where the spin effects are minor. For high BH masses, the functional form of $\Gammalc(j)$ is not obvious. In Figure \ref{fig:Gammalc_j_Gamma_lc_0}, we show the ratio $\Gammalc(j)/\Gammalc(j=0.0)$ in the spin range of $0.0\le j\le0.99$ for different BH masses $\mBH$ for a $1\,\msun,1\,\Gyr$ main-sequence star. We find that this ratio is nearly constant over the entire range of $j$ considered here for all relevant BH masses up to $10^9\msun$. For the extreme case of $\mBH=10^9\,\msun$ and $j=0.99$, the loss-cone rate is only slightly smaller than that for $j=0$, with a difference no more than 10\%.

The weak dependence of the inclination-averaged loss-cone cross-section on the BH spin is physically due to the fact that increased values of $\LTD$ and $\Lcapt$ in the retrograde orbits compensate for their decreased values in the prograde orbits in the full loss-cone case\footnote{We caution that the weak spin-dependence is not necessarily true in the empty loss-cone limit.} considered in this work. A similar conclusion was obtained by \citet{Young1977} and \citet{Kesden2012}, who found that the capture rate is nearly independent of the black hole spin $j$ if stars come from an isotropic velocity distribution. In Figure \ref{fig:Gammalc_j_Gamma_lc_0}, we also show, by a red curve, the analytical fit of $\Gammacapt(j)/\Gammacapt(j=0.0)$ for the isotropic stellar flux given by \cite{Young1977}. The analytical fit agrees with our data points of $\Gammalc(j)/\Gammalc(j=0.0)$ at $\mBH=10^9\,\msun$, because at such high BH masses, a $1\,\msun$ main-sequence star cannot be disrupted outside the horizon and hence $\Gammalc=\Gammacapt$. 

\subsubsection{Effects of stellar interior structure}

We now isolate the effects of stellar mass and age on the TDE rate by considering the maximum pericenter radii below which a star can be tidally disrupted by non-spinning BHs of different masses. The results are shown in Figure \ref{fig:rTD_mstar}.

In the hydrodynamic simulations by \cite{Ryu2020a}, it was found that the maximum pericenter for a complete tidal disruption cannot be described by a simple function of $\mstar$ but instead varies near an average value (see the right panel of their Figure 3). This complex behavior is related to the change in stellar interior structure with $\mstar$. The effect of stellar interior structure is included in our criteria for tidal disruption based on the work done by tidal forces on the fluid elements at the half-mass radius $\Rhm$ of the star.
Similar to what \cite{Ryu2020a} found, the maximum pericenter radius for tidal disruption also cannot be easily described by a simple function of $\mstar$ (for a given stellar age). One particular feature in Figure \ref{fig:rTD_mstar} is the slight decrease of the maximum pericenter when $\mstar$ goes above about $1\,\msun$. This is the consequence of the stellar density structure transitioning from that of a convective envelope (for $\mstar\lesssim 1\msun$) to a radiative envelope (for $\mstar\gtrsim 1\msun$) \citep{Kippenhahn2013}.

The interior structure of a star is also affected by its age in a number of ways. The main effects are: (1) at a given age, there is a main-sequence turn-over mass $M_{\rm max}$ and we ignore contributions to the TDE rate by post-main-sequence stars; (2) low mass stars may take up to $\sim 1\rm\, Gyr$ to contract to the main-sequence. The pre-main-sequence (PMS) contraction over time makes the star more compact and hence harder to be tidally disrupted as the star ages. To our knowledge, tidal disruption of PMS stars have not been carefully considered in the literature before. We find that young PMS stars can be disrupted by very massive BHs. For instance, a $0.2\,\msun$, $0.1\rm\,Gyr$-old PMS star can be tidally disrupted by non-spinning BHs up to $\mBH \simeq 5\times10^7\msun$ (see Table \ref{tab:mBHmax} in Appendix \ref{sec:other_tables_figures}).

\subsubsection{TDE rate fraction and the maximum BH mass}

\begin{figure}
    \centering
    \includegraphics[width=\linewidth]{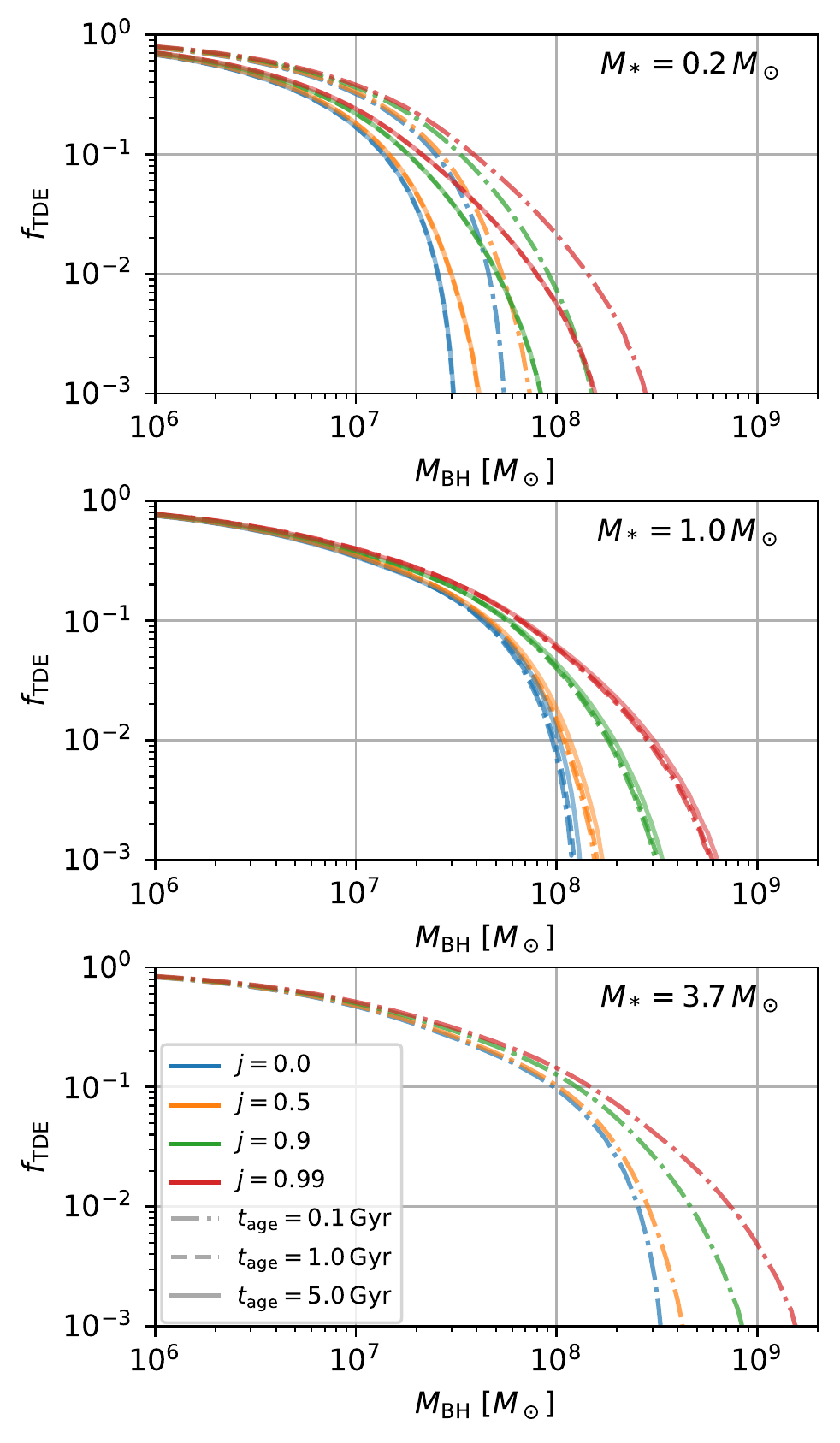}
    \caption{
    The TDE rate fraction, $\fTDE$ (the fraction of loss-cone scatterings that produce observable TDEs, eq. \ref{eq:fTDE}) as a function of BH mass and spin, for different single-star masses (different panels) and ages (different line styles).
    The line colors indicate different BH spins $j$.
    From the middle panel (for $\msun=1\,\msun$), we see that the stellar age plays a minor role for the tidal disruption of solar-like stars (only the BH spin is important).
    In the bottom panel (for $\mstar=3.7\,\msun$), only the results at $t_{\rm age}=0.1\rm\, Gyr$ are shown since the star evolves off the main-sequence long before $t_{\rm age}=1\rm\, Gyr$.
    }
    \label{fig:TDE_rate_single_star}
\end{figure}

In this subsection, we discuss the TDE fraction $\fTDE(\mBH, j)=\GammaTDE/\Gammalc$ as a function of BH mass and spin, for (single) stars of different masses and ages. The results are shown in Figure  \ref{fig:TDE_rate_single_star}.

For all stellar masses and ages, the universal trend is that $\fTDE$ drops rapidly at high BH masses, as the loss-cone scattering rates are dominated by direct captures instead of observable TDEs. The BH spin plays an important role in creating a region of large tidal forces for prograde orbits in the parameter space of $L, l_z$ that is inaccessible for non-spinning BHs. This makes it possible to, at least in principle, probe the BH spin distribution using the measured TDE rate at different BH masses.

For instance, the plummet of $\fTDE$ with the BH mass indicates the existence of a maximum BH mass $\mBHmax$ that is able to produce the observable TDE for fixed $j,\mstar,\tage$. We define $\mBHmax(j, \mstar, \tage)$ to be the critical BH mass at which $\fTDE=10^{-3}$ (the TDE rate at even higher BH masses are extremely small). The value of $\mBHmax$ for different $j,\mstar,\tage$ are shown in Figure \ref{fig:mBHmax_single_star}  and listed in Table \ref{tab:mBHmax} in Appendix \ref{sec:other_tables_figures}. We find that a solar-like star ($\mstar=1\,\msun,\tage=1.0\,\Gyr$) can be tidally disrupted by BHs up to a maximum mass that depends on the spin: $\mBHmax/(10^8\msun)\simeq 1.2$ (for $j=0$), $1.6$ ($j=0.5$), $3$ ($j=0.9$), $6$ ($j=0.99$).

Unfortunately, $\mBHmax(j, \mstar, \tage)$ also depends on the stellar mass and age in a way that is degenerate with the effects of the BH spin. Roughly speaking, more massive stars in a younger stellar population can be disrupted by more massive BHs. For instance, a high-mass young star with $\mstar=3.7\msun$ and $\tage=0.1\,\rm Gyr$ can be tidally disrupted by BHs up to $\mBHmax/(10^8\msun)\simeq 3$ (for $j=0$) and $8$ ($j=0.9)$. This means that in order to strongly constrain the BH spin distribution, one must independently constrain the mass and age distributions of the stellar population. We will return to the degeneracy in \ref{sec:age_spin_degeneracy} when considering the entire stellar population.

\begin{figure*}
    \centering
    \includegraphics[width=0.9\textwidth]{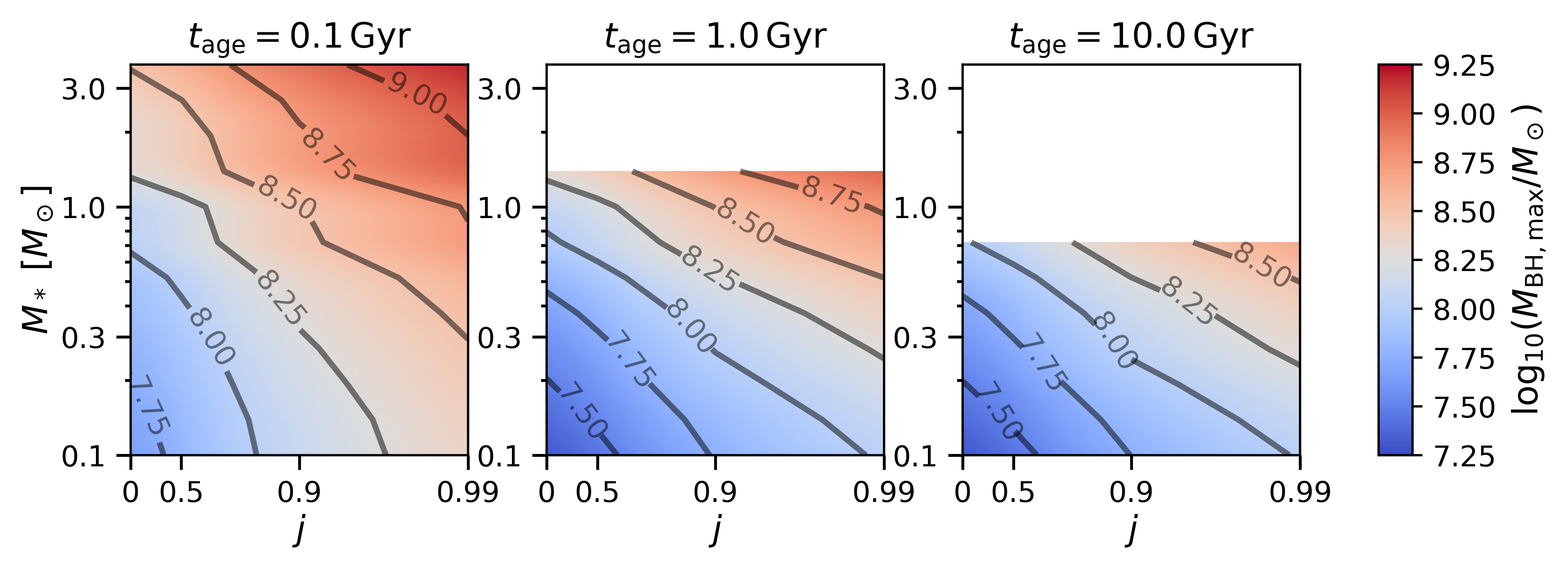}
    \caption{
    The maximum BH mass, defined by $\fTDE(\mBHmax) = 10^{-3}$, for observable TDEs for different stellar masses $M_*$ and BH spins $j$.
    Different panels show the results for different stellar ages $\tage$.
    The underlying color plot is smoothed using third-order spline interpolation for better visualization. The horizontal axis is logarithmic in $\log(1-j)$.
    }
    \label{fig:mBHmax_single_star}
\end{figure*}

\subsection{Averaging over the stellar population}\label{sec:TDE_rate_KroupaIMF}
After examining the TDE rate by stars of fixed mass and age, we now proceed to integrate the stellar populations of different masses.
We consider the simplest case where the stars surrounding the BH are formed at the same time following the Kroupa IMF.

\subsubsection{TDE rate dependence on stellar population age}\label{sec:TDE_rate_dependence_on_age}

\begin{figure*}
    \centering
    \includegraphics[width=0.9\textwidth]{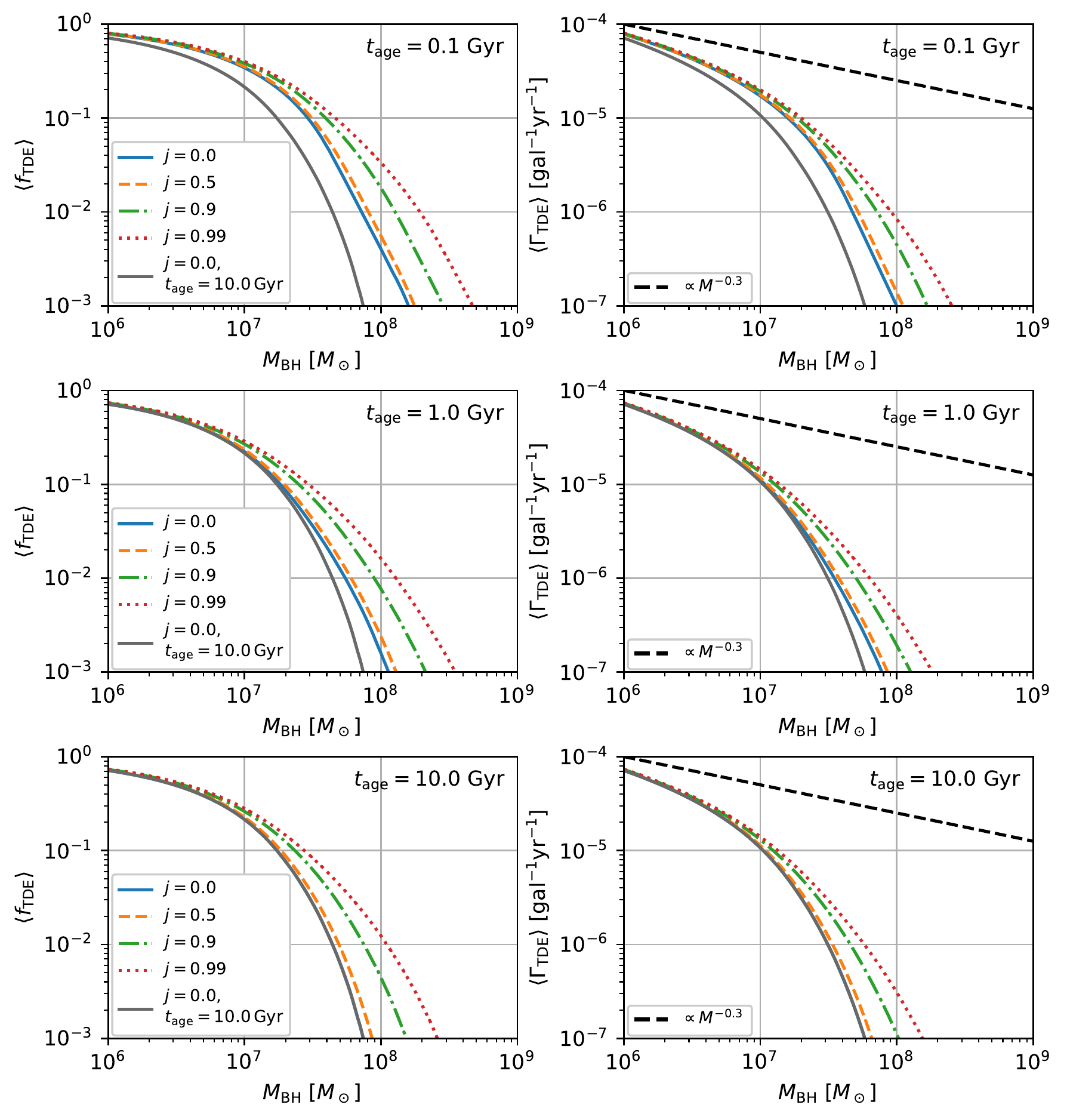}
    \caption{
    The Kroupa-population-averaged TDE rate.
    \textit{Left Panels:} $\lara{\fTDE}=\lara{\GammaTDE}/\lara{\Gammalc}$, the fraction of the TDE rate over the rate of loss-cone scatterings.
    From top to bottom, the panels show the result of different stellar population ages $\tage=0.1, \ 1, \ 10\rm\, Gyr$.
    In each panel, the curve for $j=0.0,\tage=10.0\,\Gyr$ is shown in dark-gray color to guide the eye.
    \textit{Right Panels}: Per-galaxy TDE rates given by $\lara{\GammaTDE} = \lara{\fTDE} \dot{N}$ for a power-law loss-cone scattering rate $\dot{N} = \lara{\Gammalc}=10^{-4}\lrb{\mBH/10^6\,\msun}^{-0.3}\,\rm gal^{-1}yr^{-1}$. The power-law itself is shown as the black dashed line.
    }
    \label{fig:TDE_rate_Kroupa_IMF}
\end{figure*}

The population-averaged TDE fraction $\lara{\fTDE}$, obtained based on eq. (\ref{eq:fTDE_avg}), is a function of the BH mass and spin as well as the stellar age $\tage$. This is shown in the left panels of Figure \ref{fig:TDE_rate_Kroupa_IMF}. 

Overall, $\lara{\fTDE}$ for the stellar population follows a similar trend as in Figure \ref{fig:TDE_rate_single_star} for single stars. The pre-main-sequence contraction of low-mass stars causes $\lara{\fTDE}$ to decrease at early times (before $1\,\rm Gyr$). At a fixed age $\tage$, the TDE fraction drops rapidly at high BH masses due to the dominance of direct captures.  For higher spins, TDEs can be observed up to higher BH masses. 
We define a population-averaged maximum BH mass $\lara{\mBHmax}$ by
\begin{equation}
\label{eq:mBHmax_population}
    \lara{\fTDE}(\mBH=\lara{\mBHmax})=10^{-3},
\end{equation}
which means that at higher BH masses $\mBH>\lara{\mBHmax}$, less than $10^{-3}$ of the stars that are scattered into the loss-cone would give rise to observable TDEs. The values of $\lara{\mBHmax}$ for different BH spins $j$ and stellar population ages $\tage$ are plotted in Figure \ref{fig:mBHmax_population} and listed in Table \ref{tab:mBHmax_population}  in Appendix \ref{sec:other_tables_figures}.

\begin{figure}
    \centering
    \includegraphics[width=\linewidth]{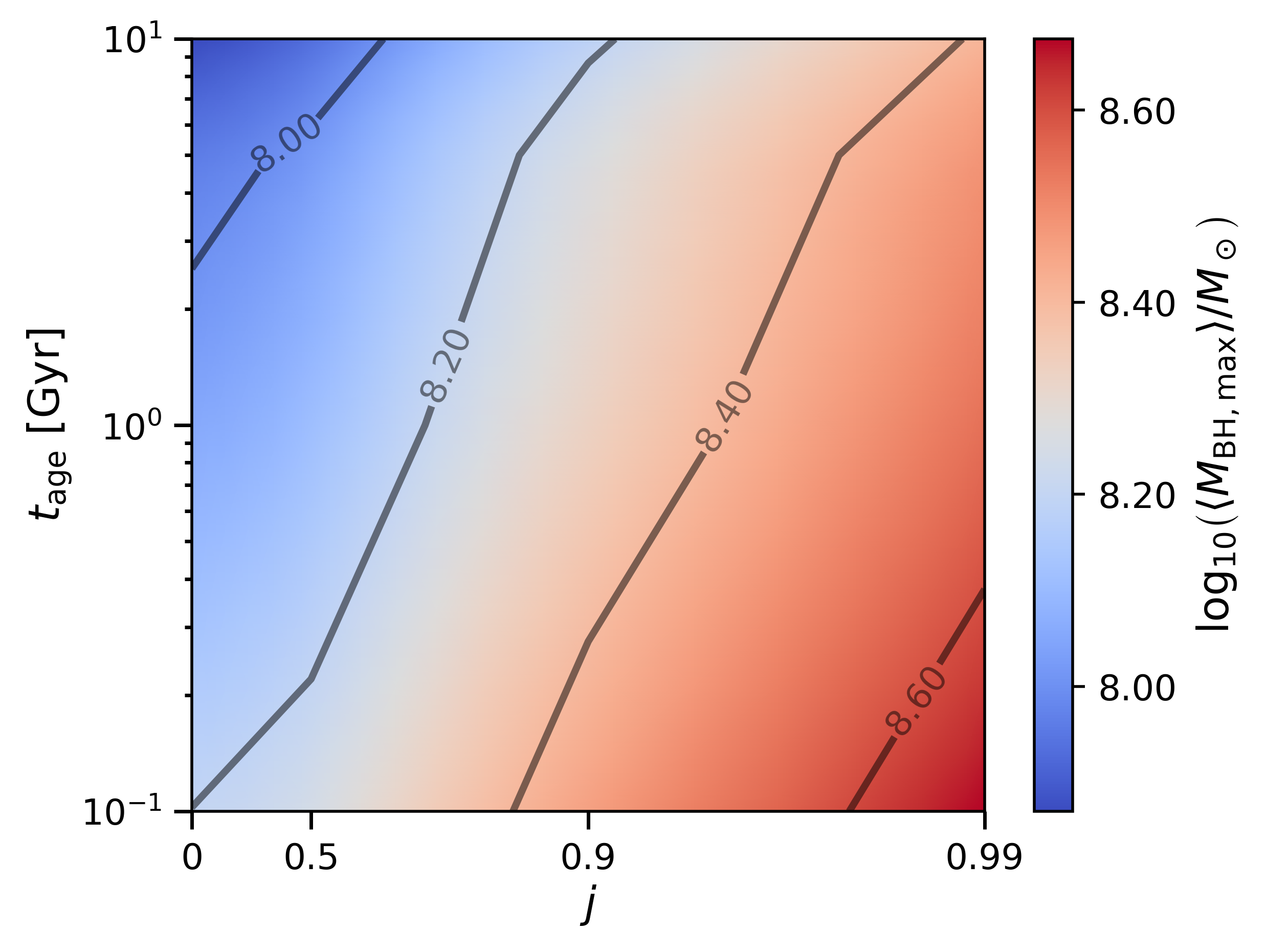}
    \caption{
    The population-averaged maximum BH mass $\lara{\mBHmax}$ (defined by eq. \ref{eq:mBHmax_population}) that can produce observable TDEs for different stellar population ages $\tage$ and BH spins $j$. The underlying color plot is smoothed using third-order spline interpolation for better visualization. The horizontal axis is logarithmic in $\log(1-j)$.
    }
    \label{fig:mBHmax_population}
\end{figure}

To convert $\lara{\fTDE}$ into the per-galaxy TDE rate, we adopt
\begin{equation}
    \lara{\GammaTDE}=\lara{\fTDE}\dot{N},
\end{equation}
where $\dot{N}=\lara{\Gammalc}$ has the following power-law form
\begin{equation}\label{eq:power-law_loss_cone_rate}
    \dot{N}(\mBH)= \lara{\Gammalc}=10^{-4}\,\lrb{\frac{\mBH}{10^6\,M_\odot}}^{-0.3}{\rm yr}^{-1}.
\end{equation}
It should be noted that our theoretical understanding of the stellar dynamics near galactic nuclei and hence the rate of loss-cone scatterings is rather limited \citep[with a number of major uncertainties, see][for a recent review]{stone20_TDE_rate_review}. However, one aspect of the functional form of $\dot{N}(\mBH)$ that is widely agreed upon is the relatively weak dependence on the BH mass. For instance, \citet{wang04_TDE_rate} found $\dot{N}\propto \mBH^{-0.25}$ by combining the $M$-$\sigma$ relation with two-body relaxation in galactic nuclei with spherically symmetric, isotropic stellar distribution function. \citet{stone16_TDE_rate} found $\dot{N}\propto \mBH^{-0.4}$ by applying the loss-cone theory to a large galaxy sample. The reason for a gradually decreasing loss-cone scattering rate with the BH mass is that the 2-body relaxation timescale near the sphere of influence is longer for galactic nuclei hosting bigger BHs, $t_{\rm rel}\propto \mBH r/\sigma\propto \mBH^2/\sigma^3$, where $r\sim G\mBH/\sigma^2$ is the radius of the BH's sphere of influence and $\sigma$ is the velocity dispersion near $r$. Making use of the $M\propto\sigma^{4.4}$ correlation \citep{kormendy13_scaling_relations}, we obtain $t_{\rm rel}\propto \mBH^{1.3}$, so the loss-cone scattering rate roughly scales as $\Gammalc\propto \mBH/t_{\rm rel}\propto \mBH^{-0.3}$.

We defer to a future work the detailed calculation of the loss-cone scattering rate, which depends on the distribution function of stars and its variation among different galaxies. In the current work, the main point is that the drop in the TDE fraction $\lara{\fTDE}$ above $\mBH\sim \mr{few}\times 10^7\,\msun$ (due to direction captures) is much steeper than the functional dependence of the loss-cone scattering rate $\dot{N}(\mBH)$. Therefore, the sharp drop of the observed TDE rate function $\lara{\GammaTDE}(\mBH)$ \citep[see Figure 3 of][]{vanvelzen18_TDE_rate_suppression} is largely due to the general relativistic effects of the event horizon. Our eq. (\ref{eq:power-law_loss_cone_rate}) is a rough representation of the weak dependence of $\dot{N}(\mBH)$, and our conclusions are only weakly affected by our choice here, although we caution that the normalization value of $10^{-4}\rm\, yr^{-1}$ in eq. (\ref{eq:power-law_loss_cone_rate}) should not be taken too seriously.



\subsubsection{Demographics in the masses of the disrupted stars}

\begin{figure}
    \centering
    \includegraphics[width=\linewidth]{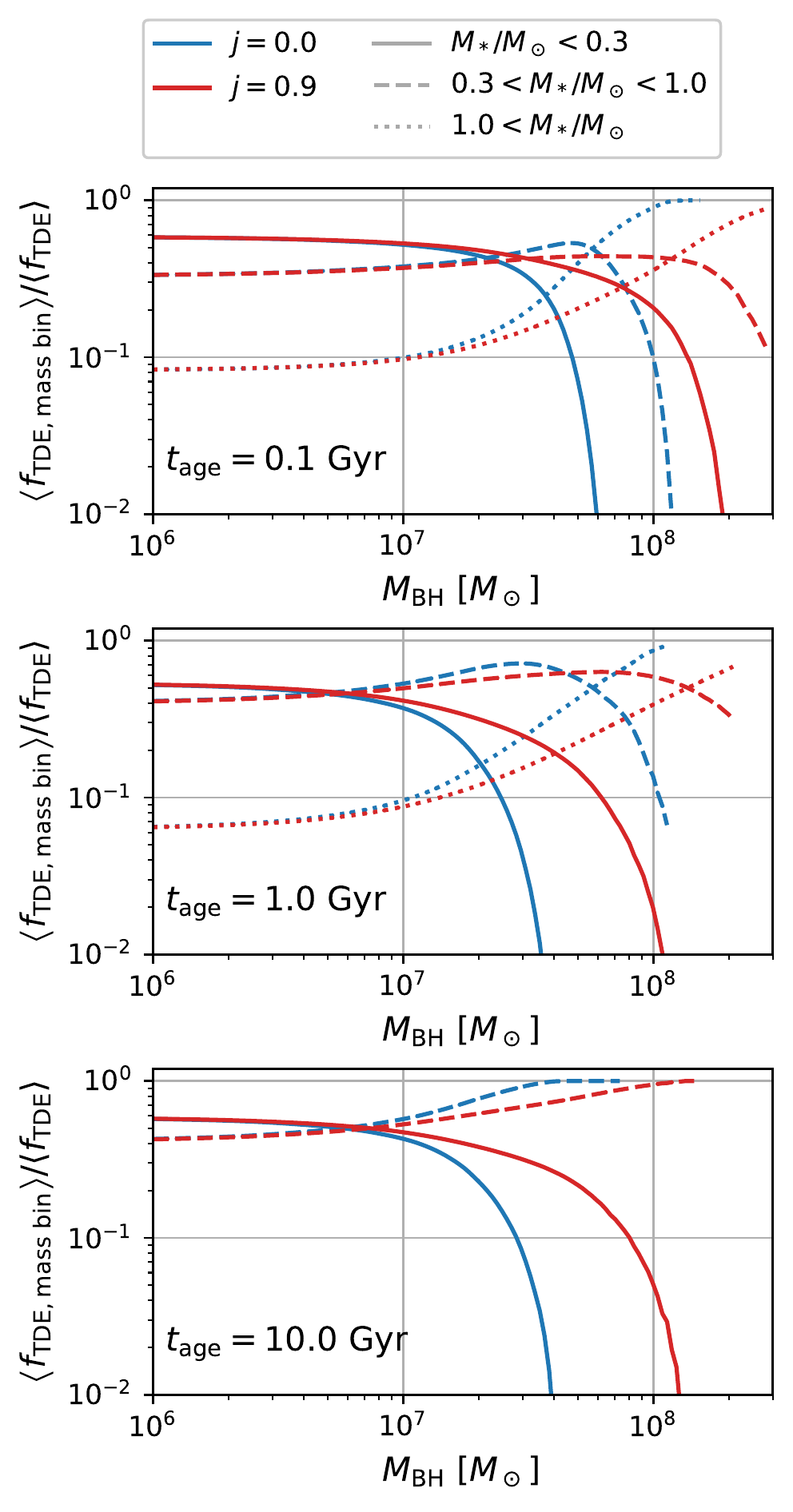}
    \caption{
    The fractional contributions to the observable TDE fraction $\lara{\fTDE}$ by stars in the three mass bins, $\mstar/\msun<0.3$ (solid), $0.3<\mstar/\msun<1.0$ (dashed), $1.0<\mstar/\msun$ (dotted lines).
    The blue and red lines represent BH spins of $j=0.0,0.9$, respectively.
    The data are only plotted if $\lara{f_{\rm TDE,mass\ bin}}>10^{-5}$ and $\lara{\fTDE}>10^{-3}$.
    }
    \label{fig:fTDE_decomposed}
\end{figure}

We then study the contributions to the TDE rate fraction $\fTDE$ from different stellar mass ranges. The results are shown in Figure \ref{fig:fTDE_decomposed}, where we divide the entire stellar population at a given age into three ranges roughly in logarithmic bins: $0.085<\mstar/\msun < 0.3$ (mostly M-dwarfs, 57.5\% of stars in the Kroupa IMF), $0.3<\mstar/\msun < 1$ (mostly K/G-types, 32.5\%), and $\mstar/\msun > 1$ (F-type and above, 10\%). Figure \ref{fig:fTDE_decomposed} shows the normalized \textit{fractional} contributions to $\lara{\fTDE}$ by these three stellar mass bins. In Figure \ref{fig:TDE_rate_decomposed_no_normalization} in Appendix \ref{sec:other_tables_figures}, we show a different version of the decomposition into three stellar mass bins, without normalizing the sum of the contributions to unity.

For low BH masses $\mBH\lesssim 10^7\msun$, the observable TDE rate is dominated by the stars in the lowest two mass bins and their contributions are comparable to each other (as dictated by the Kroupa IMF). For high BH masses $\mBH\gtrsim 10^7\msun$, the stellar demographics changes depending on the BH spin and the age of the stellar population. At the highest BH mass end, TDEs are always dominated by the highest mass stars that are still on the main-sequence (for a given age). For instance, at an age of $\tage=0.1\rm\, Gyr$, the highest stellar mass bin dominates the TDE rate at $\mBH\gtrsim 10^8\msun$ for both spins $j=0$ and $0.9$. At older ages $\tage\gtrsim 1\rm\, Gyr$, stars more massive than about $2.5(\tage/\mr{Gyr})^{-0.4}\msun$ have evolved off the main-sequence and hence the contribution to the observable TDE rate by the highest stellar mass bin decreases rapidly with stellar age. 


\subsubsection{Spin-Age Degeneracy (SAD)}\label{sec:age_spin_degeneracy}

As can be clearly seen from Figure \ref{fig:TDE_rate_Kroupa_IMF}, \ref{fig:mBHmax_population} and Table \ref{tab:mBHmax_population} in Appendix \ref{sec:other_tables_figures}, there is a degeneracy between the age of the stellar population and the BH spin, because BHs with $\mBH\gtrsim 10^8\msun$ can produce observable (by ``observable'' we mean $\lara{\fTDE}>10^{-3}$) TDEs provided that the stellar population is young $\tage\lesssim 1\rm\, Gyr$ or the spin is high $a\gtrsim 0.5$ (or a combination of these two). A possible example of such TDEs is ASASSN-15lh, where the inferred mass of the BH at the nucleus of the host galaxy is $\mBH\gtrsim2\times10^8\,\msun$ \citep{leloudas16_AS15lh}. At such high BH masses, TDEs are necessarily limited to the region close to the horizon. High spins make it possible for low-mass stars in prograde orbits to reach closer to the BH and experience stronger tidal forces. Meanwhile, a younger age means that short-lived high-mass stars, which are easier to tidally disrupt, can give rise to observable TDEs. Hereafter, we call this spin-age degeneracy (SAD).

To further explore the competition between the spin and the age of the stellar population, we show the population-averaged TDE fraction $\lara{\fTDE}$ for two selected BH masses, $\mBH=10^7,10^{7.7}\,\msun$, in Figure \ref{fig:TDE_rate_j_Kroupa_IMF} and in Table \ref{tab:fTDE} in Appendix \ref{sec:other_tables_figures}. We see that for $\mBH=10^7\,\msun$ (or lower BH masses), the effects of spin and stellar age are both mild. The main change in $\lara{\fTDE}(\mBH=10^7\,\msun)$ occurs between $\tage=0.1\,\Gyr$ and $1.0\,\Gyr$, mainly due to the contraction of low-mass pre-main-sequence stars. For $\mBH=10^7\,\msun$, TDE fraction at $\tage\gtrsim 1\rm\, Gyr$ does not depend on the age of the stellar population and only slightly increases with $j$. However, for $\mBH= 10^{7.7}\,\msun$ (or higher BH masses), the effects of $j$ and $\tage$ are both prominent. For instance, a young stellar population at $\tage=0.1\rm\, Gyr$ produces a much larger TDE fraction than an old population at $\tage=10\rm\,Gyr$ by a factor of 3 to 5 (depending on the spin). A high BH spin of $j=0.9$ also produces a larger TDE fraction than the case of zero spin by a factor of 2 to 5 (depending on the stellar age).

Therefore, we emphasize the importance of including the effects of stellar population ages when using the TDE rates to constrain the BH spin distribution.

\begin{figure}
    \centering
    \includegraphics[width=\linewidth]{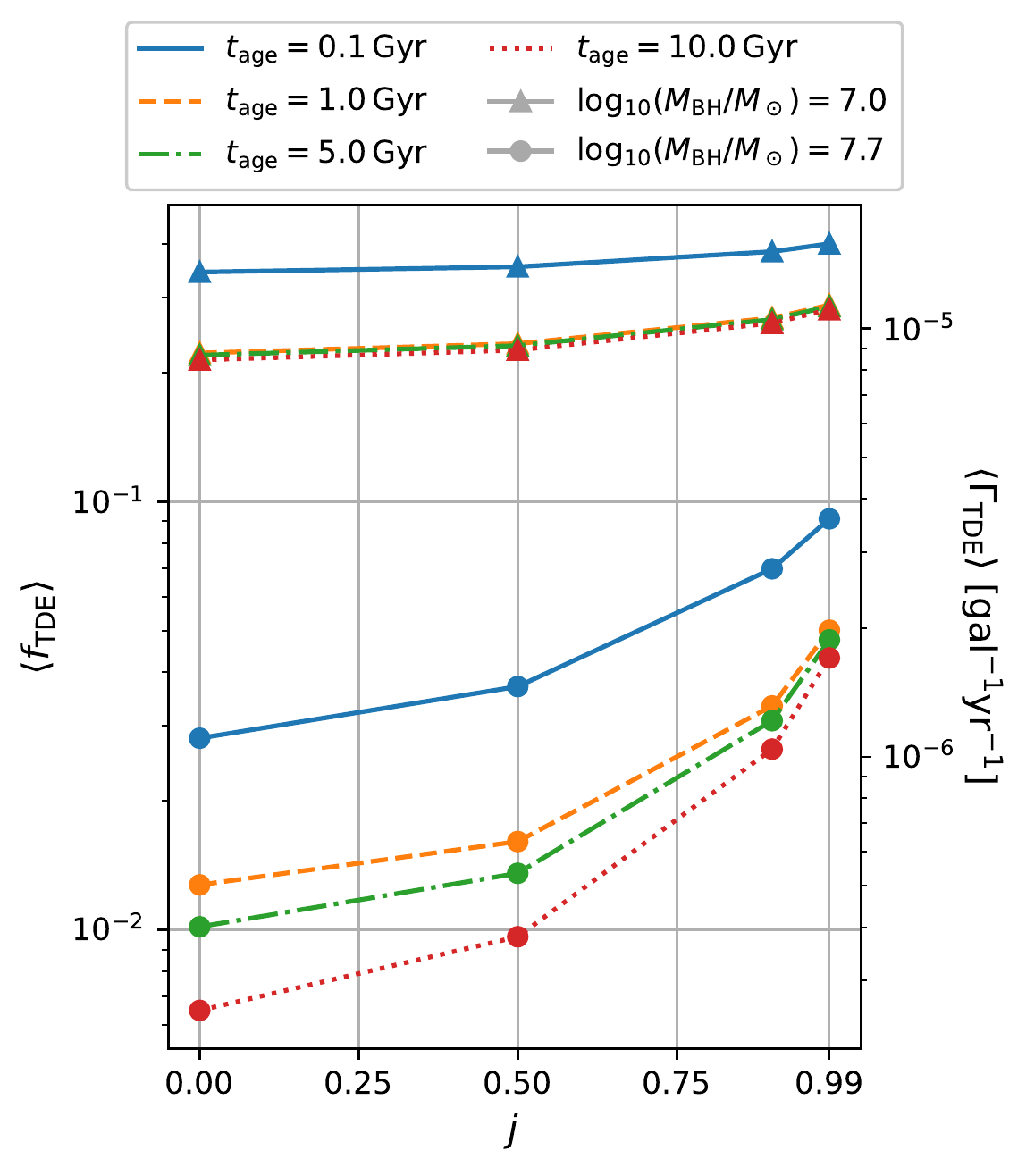}
    \caption{
    Population-averaged TDE fraction $\lara{\fTDE}$ for different ages $\tage$ and spins $j$, at two selected BH masses, $\log_{10}(\mBH/\msun)=7.0$ (triangles) and $7.7$ (circles).
    The right vertical axis uses the loss-cone scattering rate $\dot{N}=10^{-4}\lrb{\mBH/10^6\,\msun}^{-0.3}\,\rm gal^{-1}yr^{-1}$ evaluated at the geometric-mean BH mass, $\log_{10}(\mBH/\msun)=7.35$, to convert $\lara{\fTDE}$ into the observable TDE rate $\lara{\GammaTDE}$. The data used in the figure are listed in Table \ref{tab:fTDE}.
    }
    \label{fig:TDE_rate_j_Kroupa_IMF}
\end{figure}

\section{Discussion}\label{sec:discussion}


In this section, we discuss the limitations and caveats in this work and how future works can improve upon our calculations.

(1) Our relativistic criteria for tidal disruption are different from the previous studies that only consider the maximum strength of the tidal forces at the pericenter \citep{Kesden2012,Coughlin2022a,Coughlin2022b}. By integrating the geodesic deviation equation (eq. \ref{eq:geodesic-deviation}), our criteria explicitly calculate the work done by the tidal forces during the pericenter passage under the framework of the frozen-in approximation. The consideration of the work done by the tidal forces, as opposed to only the maximum strength, is important in the tidal disruption condition (see Figure \ref{fig:Ryu_et_al_compare}).

In \citet{Ryu2020b}, it is argued that the physical tidal disruption radius can be modeled by equating the tidal force at the pericenter to the stellar gravity scaled by a fixed constant (see their eq. 12 and Figure 10). A similar treatment is also proposed by \citet{Coughlin2022b}. Even though \citep{Ryu2020b} has demonstrated the applicability of this argument for low BH masses ($10^6\,\msun$), it is unlikely to hold at higher BH masses, where the TDE is increasingly relativistic. If the disruption criteria via the maximum tidal forces were to hold, we would expect the hydrodynamic simulation results of \cite{Ryu2020d} to trace out the contours of maximum tidal forces\footnote{Consequently, the suppression of the TDE rate at high BH masses would be much steeper \citep[see Figure 6 of][]{Coughlin2022a} compared to our results.} in Figure \ref{fig:Ryu_et_al_compare}. However, even with the explicit form of the relativistic tidal force, the contours of maximum tidal forces (dashed lines) fail to reproduce the results of hydrodynamic simulations. This shows that there is no single scaling factor to predict the disruption of a given star across all BH masses; instead, the work done by tidal forces must be taken into consideration.

We caution that our criteria, based on the combination of the work done by tidal forces and the relativistic Roche lobe condition, are only approximate and need to be further verified by future hydrodynamic simulations in Kerr spacetime. Nevertheless, this is the first attempt to incorporate full general relativity and stellar density structure to predict the TDE rate. This allows us to make direct comparison to the observed TDE rate from on-going and future surveys.

(2) Our model cannot predict the loss-cone scattering rate in a galaxy, which involves the distribution function of stars, their mutual interactions, as well as interactions with other massive objects in galactic nuclei. Instead, we assume a loss-cone scattering rate $\dot{N}$, which can in principle be obtained from the loss-cone theory \citep{Merritt2013}, and then calculate the fraction of these stars that produce observable TDEs $\fTDE$. The \textit{independent} combination of the loss-cone theory with our relativistic tidal disruption criteria, however, has some important caveats.

The first caveat is that the loss-cone theory must be fully general relativistic, which is non-trivial\footnote{In the empty loss-cone limit and for sufficiently high BH masses, the spin of the BH makes the stellar distribution function non-spherical and anisotropic, since retrograde orbits have higher direct capture cross-sections which depends on the orbital inclination wrt. the BH spin axis. This makes the general relativistic loss-cone theory non-trivial (on top of the large uncertainties in the velocity and density distributions of stars).}. The second caveat is the nature of the loss cone. Our assumption of an isotropic velocity distribution of stars can be effectively viewed as the full loss-cone regime. This assumption, however, is highly idealized, as the relaxation timescale for stars near high-mass BHs is much longer. Taking stars near the sphere of influence $r\sim G\mBH/\sigma^2$ as an example (where $\sigma$ is the velocity dispersion near $r$), the angular momentum diffusion timescale for highly eccentric orbits with pericenter radii $\rp\ll r$ is roughly given by $t_{\rm J} \sim (\rp/r)t_{\rm rel}$, where $t_{\rm rel}$ is the 2-body relaxation timescale near $r$. The orbital period is $P(r)\sim r/\sigma$, so we obtain the following scaling $t_{\rm J}/P \propto \rp t_{\rm rel}\sigma^5/\mBH^2$. For high-mass BHs, we take $\rp \propto \rg \propto \mBH$ and $t_{\rm rel}\propto \mBH^{1.3}$ obtained based on the $M$-$\sigma$ correlation of $M\propto \sigma^{4.4}$ (see \S \ref{sec:TDE_rate_dependence_on_age}), and then the ratio between the two timescales scales as $t_{\rm J}/P\propto \mBH^{1.4}$. Stars are in the empty (or full) loss-cone regime when $t_{\rm J}\gg P$ (or $t_{\rm J}\ll P$).
We see that it is likely that the stars in the nuclei of very massive BHs are in the empty loss-cone regime, as argued by \citet{Merritt2013}. If this is the case, our full-loss-cone assumption then leads to an underestimate of the TDE fraction $\fTDE$, meaning that our predicted TDE rate should be considered as lower limits when compared with observations.
The investigation of the effects of the more realistic loss cone near spinning BHs is deferred to a future work.

(3) Throughout this work, we take 50\% mass loss as a representative criterion for an observable TDE. This assumption is partly due to our poor understanding of the (optical and X-ray) emission mechanisms of TDEs. More realistically, partial TDEs with smaller mass loss fractions may be observable from nearby galaxies. In the future when better understanding of the TDE emission (especially in the optical band) is available, one might revise our strict cut of 50\% mass loss to other values. For instance, for the case of 30\% mass loss, our tidal disruption criteria would then need to be modified to consider the fluid elements at a radius of $R_{30}$ (for exterior mass of $0.3\mstar$) and the maximum differential velocity $\Delta v_{\rm max}$ due to tidal forces needs to be compared with the virial velocity of $v_{\rm vir} = \sqrt{GM_*(1-30\%)/R_{30}}$. Because the fractional mass loss from the star is a very steep function of the pericenter radius \citep{guillochon13_TDE_hydro, Ryu2020a}, we expect that our results will only be weakly affected by this aspect of uncertainty. We leave to a future work to explore in detail other fractions of mass loss from the star.

(4) We have assumed, for simplicity, a single stellar population as given by the Kroupa IMF. There is evidence that the IMF near our own Galactic Center is more top-heavy \citep{lu13_top_heavy_IMF}. Such top-heavy IMFs will increase the TDE rate from the most massive BHs as compared to our predictions. Recent observations showed that the post-starburst galaxies, despite their rarity, are over-represented in the detected TDE sample \citep{French2020,Hammerstein2021}. Moreover, \cite{Bortolas2022} studied the dynamics of nuclear star clusters with IMFs with various degrees of top-heaviness and found that the loss-cone scattering rate is strongly enhanced due to mass segregation in the early evolution of the cluster at $\tage\lesssim 1\rm\, Gyr$ as compared to an old cluster at $\tage=10\rm\, Gyr$. These previous works, together with our findings, demonstrate that the influence of the stellar population age on the observed TDE rate cannot be ignored --- the stellar population age, if unconstrained, will severely compromise the utility of TDEs as a probe of BH spin distribution (see \ref{sec:age_spin_degeneracy}). Self-consistent calculations of the TDE rate must consider the \textit{current} stellar mass function and the dynamics in galactic nuclei in the framework of the relativistic loss-cone theory for spinning BHs.

(5) Finally, we have assume that a star that is tidally disrupted outside the horizon would produce an electromagnetically bright and detectable signal. An important point is that, for the extreme cases near the highest mass BHs, about half of the tidally stripped debris would plunge into the BH and the other half becomes unbound. The question is whether the bound (but plunging) debris would produce bright emission before entering the event horizon. Recently, \citet{Ryu22_extremely_relativistic_TDEs} carried out hydrodynamic simulation of a TDE with stellar pericenter distance of $\rp=4.02\,\rg$ for a non-spinning BH. They found that, although the bound debris are largely in plunging orbits (and certainly do not form a rotationally supported accretion disk), internal shocks form as the material fall towards the BH due to apsidal precession and that the radiative efficiency (in terms of rest mass) is of the order a few percent. Therefore, it is possible that even extremely relativistic TDEs produce detectable emission.

\section{Summary}\label{sec:summary}

In this paper, we quantify the suppression of the observable TDE rate due to direct captures of stars by the event horizon of spinning BHs, improving upon the work of \citet{Kesden2012}.

We first generalize the commonly adopted frozen-in approximation from the Newtonian limit to the general relativistic case, including the effects of stellar interior structure. This is achieved by integrating the equation of motion according to the tidal tensor in the comoving frame of the center of mass of the star on a Kerr geodesic. Our integration starts when the star first enter the sphere of radius $r_0$ from the BH and ends when it exits from the sphere. By uniformly sampling the fluid elements on the sphere at half-mass radius $\Rhm$ from the stellar center, we obtain the maximum velocity $\Delta v(r_0)$ achieved by all these fluid elements in the comoving frame of the star's center of mass. The choice of half-mass radius is motivated by our consideration of 50\% mass loss from the star as the threshold for bright TDEs (partial TDEs with a much smaller fractional mass loss would be fainter and more difficult to detect). Then, by varying the initial radius $r_0$, we obtain the maximum velocity $\Delta v_{\rm max} = \max_{\{\mr{all\ } r_0\}}[\Delta v(r_0)]$, which represents the maximum possible work done by tidal forces on any of the fluid elements at the star's half-mass radius. We further restrict the initial radius $r_0$ by requiring that an unperturbed star fills up the relativistic Roche lobe at all radii $r < r_0$, because otherwise there will be no mass loss from the star. Finally, if the maximum differential velocity between the fluid element and the star's center of mass exceeds the virial velocity at half-mass radius $v_{\rm vir} = \sqrt{G\mstar/(2\Rhm)}$, we infer that the star will lost more than 50\% of its mass and that there will be a bright, detectable TDE. The above tidal disruption criteria are in good agreement with the results from the relativistic hydrodynamic simulations carried out by \citet{Ryu2020d} (see Figure \ref{fig:Ryu_et_al_compare} for a comparison).

The next step is to consider the angular momentum distribution of stars that are scattered into the loss-cone. We consider the full loss-cone case and calculate the ratio between the rates of TDEs and direct captures, $\GammaTDE/\Gammacapt$, for a given star (the ratio is directly given by the cross-sections of TDEs and direct captures). The observable TDE fraction is then defined as $\fTDE = \GammaTDE/(\GammaTDE + \Gammacapt)$, where the denominator is the total rate of loss-cone scatterings $\Gammalc = \GammaTDE + \Gammacapt$. We then consider a stellar population as given by the Kroupa IMF at different ages. By integrating over the stellar mass distribution at a given age of the population, we calculate the population-averaged observable TDE fraction $\lara{\fTDE} = \lara{\GammaTDE}/\lara{\Gammalc}$, which depends on the BH mass $\mBH$ and dimensionless spin $j$, and the age of the stellar population $\tage$.

The population-averaged TDE rate per galaxy is then given by $\lara{\GammaTDE} = \lara{\fTDE}(\mBH, j, \tage) \times \lara{\Gammalc}$, where $\lara{\Gammalc}$ is the population-averaged rate of loss-cone scatterings. To make a direct comparison with observational measurements of the TDE rate as a function of the BH mass (provided that $\mBH$ can be inferred from e.g., the $M$-$\sigma$ correlation), we further need the loss-cone scattering rate $\lara{\Gammalc}$. Despite uncertainties in the distribution function of stars near galactic nuclei, it is theoretically expected that $\lara{\Gammalc}$ only depends weakly on the BH mass across different galaxies \citep[e.g.,][]{wang04_TDE_rate, Merritt2013, stone16_TDE_rate}. Thus, the observed TDE rate is mainly sensitive to the TDE fraction $\lara{\fTDE}(\mBH, j, \tage)$, and this makes it possible to achieve an important goal of the TDE community --- to constrain the spin distribution of dormant BHs.

The results in this work make the first step towards this goal. However, we find that a serious hurdle to overcome is the spin-age degeneracy (SAD), which means that either a young stellar population ($\tage\lesssim 1\rm\, Gyr$) or a high BH spin ($a\gtrsim 0.5$), or a combination of these two factors, can extend the mass function of TDE-hosting BHs significantly above $10^8\,\msun$. A possible example of such TDEs is ASASSN-15lh, where the inferred mass of the hosting BH is $\mBH\gtrsim 2\times10^{8}\,\msun$ \citep{leloudas16_AS15lh}. 

To break this degeneracy with complimentary information, we suggest the following strategies:
\begin{itemize}
    \item Make use of the electromagnetic signals from the TDE to independently constrain the mass of the disrupted star \citep[e.g.,][]{Mockler22_stellar_mass_constraints}. However, this requires a significant improvement in our understanding of the multi-wavelength emission mechanisms of TDEs.
    \item Systematically search for signatures (in the X-ray and radio bands) of relativistic jets from TDEs hosted by the most massive BHs. According to the \citet{Blandford77_BHjet} mechanism, relativistic jets are most likely associated with high BH spins.
    \item Use high spatial resolution imaging or spectroscopy to obtain the spectral energy distribution (SED) or spectrum of the stars near the TDE-hosting galactic nuclei, after the TDE has faded away. However, the TDE emission may over-shine the stellar emission for many decades in the UV bands \citep{vanVelzen19_late_time_HST_UV}.
\end{itemize}


\section*{Acknowledgments}
HTH would like to thank the financial support from Department of Physics, The Chinese University of Hong Kong for this research. 

\section*{Data Availability}
The data underlying this article will be shared on reasonable request to the corresponding author.

{\small
\bibliographystyle{mnras}
\bibliography{refs}
}

\appendix
\section{Tidal Tensor}\label{sec:tidal_tensor}
Here we provide the full expression of tidal tensor in Kerr spacetime:
\allowdisplaybreaks
\begin{align}
C_{11} = & \lrsb{1-3\frac{ST(r^{2}-a^{2}\cos^{2}\theta)}{K\Sigma^{2}}\cos^{2}\Psi}I_{1}\nonumber\\
         & +6ar\cos\theta\frac{ST}{K\Sigma^{2}}\cos^{2}\Psi I_{2},\\
C_{12} = & \lrsb{-ar\cos\theta(S+T)I_{1}+(a^{2}\cos^{2}\theta S-r^{2}T)I_{2}}\nonumber\\
         & \times3\frac{\sqrt{ST}}{K\Sigma^{2}}\cos\Psi,\\
C_{13} = & \lrsb{(a^{2}\cos^{2}\theta-r^{2})I_{1}+2ar\cos\theta I_{2}}3\frac{ST}{K\Sigma^{2}}\cos\Psi\sin\Psi,\\
C_{22} = & \lrb{1+3\frac{r^{2}T^{2}-a^{2}\cos^{2}\theta S^{2}}{K\Sigma^{2}}}I_{1}-6ar\cos\theta\frac{ST}{K\Sigma^{2}}I_{2},\\
C_{23} = & \lrsb{-ar\cos\theta(S+T)I_{1}+(a^{2}\cos^{2}\theta S-r^{2}T)I_{2}}\nonumber\\
         & \times3\frac{\sqrt{ST}}{K\Sigma^{2}}\sin\Psi,\\
C_{33} = & \lrsb{1-3\frac{ST(r^{2}-a^{2}\cos^{2}\theta)}{K\Sigma^{2}}\sin^{2}\Psi}I_{1}\nonumber\\
         & +6ar\cos\theta\frac{ST}{K\Sigma^{2}}\sin^{2}\Psi I_{2},
\end{align}
where
\begin{align}
S & =K+r^{2},\\
T & =K-a^{2}\cos^{2}\theta,\\
I_{1} & =\frac{\mBH r}{\Sigma^{3}}(r^{2}-3a^{2}\cos^{2}\theta),\\
I_{2} & =\frac{\mBH a\cos\theta}{\Sigma^{3}}(3r^{2}-a^{2}\cos^{2}\theta),
\end{align}
and $K$ is related to the Carter constant $Q$ by
\begin{equation}
    K=Q+\lrb{L_z-aE}^2.
\end{equation}
The expression of $C_{21},C_{31},C_{32}$ follows directly from the symmetry of the tidal tensor.
The angle $\Psi$ in the above expressions is the rotation angle that ensures the basis of the coordinates is parallel transported.
The value of $\Psi$ changes along the geodesic according to
\begin{equation}
    \frac{d\Psi}{d\tau}=\frac{\sqrt{K}}{\Sigma}\lrsb{\frac{(r^{2}+a^{2})E-aL_{z}}{K+r^{2}}+a\frac{L_{z}-aE\sin^{2}\theta}{K-a^{2}\cos^{2}\theta}}.
\end{equation}
We refer the readers to \cite{Marck1983} for a complete derivation behind those expressions.

\section{Relativistic Roche lobe}\label{sec:Roche_lobe}
Consider a star subjected to the forces from a tidal tensor $C_{ij}$.
The tidal force $F^i$ at a given position can be written in Fermi Normal Coordinates as
\begin{equation}
    F^i=-C_{ij}\chi^j,
\end{equation}
where $\boldsymbol{\chi}$ is the displacement 3-vector from the star center.
The symmetric tidal tensor allows the definition of the tidal potential
\begin{equation}
    \Phi_{\rm tide}=\frac{1}{2}C_{ij}\chi^i\chi^j,
\end{equation}
such that
\begin{equation}
    \vec{F}=-\nabla\Phi_{\rm tide}.
\end{equation}
Furthermore, the symmetry of the tidal tensor ensures the existence of a rotated coordinate $\Tilde{\boldsymbol{\chi}}$ in which the tidal tensor is diagonal:
\begin{align}
    \Tilde{C}_{ij} = & \begin{pmatrix} \xi_1 & 0 & 0 \\ 0 & \xi_2 & 0 \\ 0 & 0 & \xi_3 \end{pmatrix}, \\
    \Phi_{\rm tide}= & \frac{1}{2}\sum_{i=1,2,3}\xi_i(\Tilde{\chi}^i)^2,
\end{align}
where $\xi_i$ are the eigenvalues of the tidal tensor.
Let $\xi_1$ be the smallest eigenvalue among $\xi_i$ and
$\xi_1$ is guaranteed to be negative because the tidal tensor is traceless.

The total gravitational potential experienced by the star $\Phi_{\rm total}$ is the sum of $\Phi_{\rm tide}$ and its own self-gravity $\Phi_{\rm star}$:
\begin{equation}
    \Phi_{\rm total}=\Phi_{\rm tide}+\Phi_{\rm star}.
\end{equation}
The tidal forces will deform the star and $\Phi_{\rm star}$ will in general deviate from the potential of a perfect sphere.
Nonetheless, we assume that prior to reaching the initial radius $r_0$, the deformation of the star is small and its potential can be well-approximated as spherical.
In the natural units $G=\mBH=c=1$,
\begin{equation}
    \Phi_{\rm star}=-\frac{\mstar/\mBH}{\abs{\Tilde{\boldsymbol{\chi}}}}.
\end{equation}
Unlike the binary system, $\Phi_{\rm total}$ here does not contain the potential of the centrifugal force as we assume the star to be non-rotating.

The inclusion of $\Phi_{\rm tide}$ introduces a critical value of potential $\Phi_{\rm crit}$, above which the equipotential surface is no longer closed within the inner Lagrangian point.
The equipotential surface of $\Phi_{\rm crit}$ thus marks the maximum extent that a star in hydrostatic equilibrium can reach before starting to lose mass.
We refer to the equipotential surface $\Phi_{\rm crit}$ as the ``Roche lobe'', although it differs from the classical Roche lobe in a circular binary system where the donor star is assumed to be in synchronous rotation, because here we assume the star to be non-rotating (and hence the potential $\Phi_{\rm tot}$ does not include the centrifugal term).

The value of $\Phi_{\rm crit}$ is determined by the saddle point in $\Phi_{\rm total}$, which can be obtained by solving $\nabla\Phi_{\rm total}=0$ and is given by
\begin{equation}
    \Phi_{\rm crit}=-\frac{3}{2}(-\xi_1)^{1/3}(\mstar/\mBH)^{2/3}.
\end{equation}
The volume of the Roche lobe $\VRL$ can be written as
\begin{equation}
    \VRL=\int_0^{\pi}\int_0^{2\pi}\frac{\tr^3(\ttheta,\tphi)}{3}d\tphi d\ttheta,
\end{equation}
where $\tr,\ttheta,\tphi$ are the spherical coordinates of the frame $\Tilde{\chi}$:
\begin{align}
    \Tilde{\chi}^1 = & \tr\sin\ttheta\cos\tphi, \\
    \Tilde{\chi}^2 = & \tr\sin\ttheta\sin\tphi, \\
    \Tilde{\chi}^3 = & \tr\cos\ttheta.
\end{align}
The value of $\tr$ for a given set of $\ttheta,\tphi$ is determined by
\begin{align}
\label{eq:Roche_lob_rtilde}
    \Phi_{\rm crit} = & \frac{1}{2}\sum_{i=1,2,3}\xi_i(\Tilde{\chi}^i)^2-\frac{\mstar/\mBH}{\abs{\Tilde{\boldsymbol{\chi}}}}\nonumber\\
                    = & \frac{1}{2} \tr^2\lrb{\xi_1\sin^2\ttheta\cos^2\tphi+\xi_2\sin^2\ttheta\sin^2\tphi+\xi_3\cos^2\ttheta}\nonumber\\
                      & - \frac{\mstar/\mBH}{\tr}.
\end{align}

An additional property of the Roche lobe is gained by observing that the above equation (Eq. \ref{eq:Roche_lob_rtilde}) can be rewritten in a scaled radius $\tr^{\prime}=\tr(\mstar/\mBH)^{-1/3}$ and has the form
\begin{multline}
    -\frac{3}{2}(-\xi_1)^{1/3}=\\
    \frac{1}{2} (\tr^{\prime})^2\lrb{\xi_1\sin^2\ttheta\cos^2\tphi+\xi_2\sin^2\ttheta\sin^2\tphi+\xi_3\cos^2\ttheta}-\frac{1}{\tr'}.
\end{multline}
The value of $\tr^{\prime}$ is therefore independent of the stellar mass and completely given by the eigen values $\xi_i$ and the angles $\ttheta,\tphi$.
This implies that the volume of the Roche Lobe $\VRL$ is proportional to $\mstar$.

\section{Differential rate}\label{sec:differential_rate}
\begin{figure}
    \centering
    \includegraphics[width=0.9\linewidth]{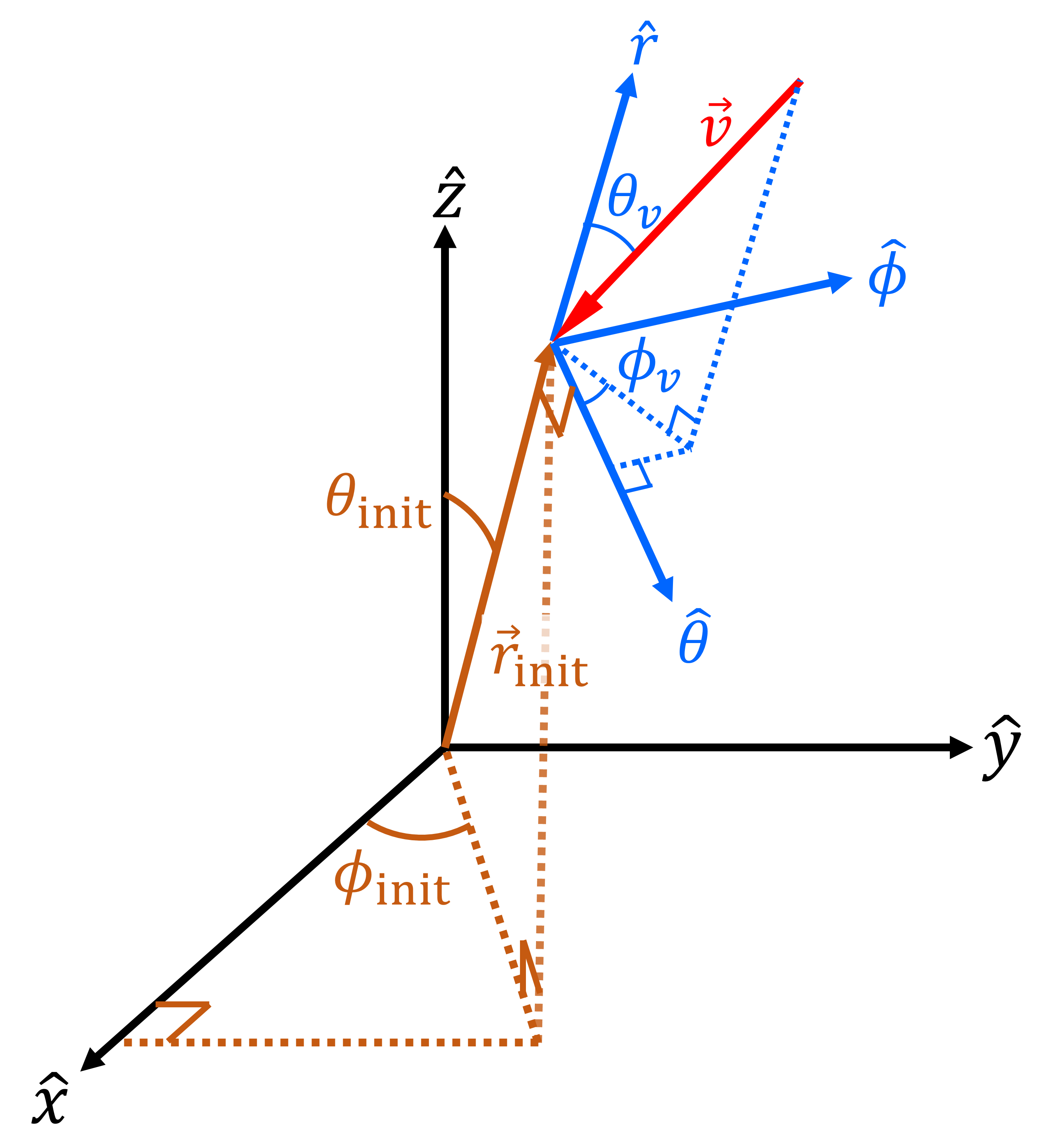}
    \caption{
    The initial movement of the star described by the angles $\thetainit,\phiinit,\theta_v,\phi_v$. 
    The SMBH is located at the origin with the spin axis in $z$ direction.
    The brown vector indicates the initial position of the star $\vec{r}_{\rm init}$ and the red vector indicates the initial velocity of the star $\vec{v}$, which are described by black, local coordinates $\hat{x},\hat{y},\hat{z}$, and blue, spherical coordinates $\hat{r},\hat{\theta},\hat{\phi}$, respectively.
    Notice that the velocity of the star (red vector) is pointing inward as we only consider the star that can approach the vicinity of the SMBH.
    }
    \label{fig:TDE_coordinates}
\end{figure}

Consider the stars being shot from the surface of a sphere with radius $\rinit$ towards the SMBH located at the origin.
The stellar number density $n$ is uniform on the sphere and the velocity of the stars $v$ is fixed.
The direction of stellar velocity is assumed to be isotropically distributed.
The initial motion of a star on the sphere can be described by a set of angles $\thetainit,\phiinit,\theta_v,\phi_v$ and is illustrated in Figure \ref{fig:TDE_coordinates}:
$\thetainit$ and $\phiinit$ describe the initial position of the star in the Boyer-Lindquist coordinates of the Kerr metric (eq. \ref{eq:BL_coor}).
At the large distance $\rinit\gg\rg$, $\thetainit,\phiinit$ can be viewed as the polar and azimuthal angles of the spherical coordinates.
In the local frame $(\hat{r},\hat{\theta},\hat{\phi})$ of a given star at its initial position, $\theta_v$ and $\phi_v$ are the angles that the velocity of the star makes with the inward radial direction $-\hat{r}$ and $\hat{\theta}$ in the plane perpendicular to $\hat{r}$, respectively.
In terms of $\thetainit,\theta_v,\phi_v$, the differential rate of the stars can be expressed as
\begin{equation}
    \frac{\partial^3\Gamma}{\partial\thetainit\partial\theta_v\partial\phi_v}=\frac{1}{2}r_0^2nv\sin\thetainit\sin\theta_v\cos\theta_v.
\end{equation}
Note that we only consider the stars entering the sphere, with $\theta_v\in[0,\pi/2]$.

The total specific angular momentum $L$ and the fractional angular momentum in the direction of the black hole spin axis $l_z=L_z/L$ are related to the above angles by
\begin{align}
    L =   & r_0v\sin\theta_v,\\
    l_z = & -\sin\thetainit\sin\theta_v.
\end{align}
Using the change of variables, the differential rate can be expressed in terms of $L,l_z,\thetainit$ as
\begin{equation}
    \frac{\partial^3\Gamma}{\partial\thetainit\partial L\partial l_z}=\frac{n}{v}\frac{L}{\sqrt{1-\lrb{l_z/\sin\thetainit}^2}}.
\end{equation}
For a given $l_z$, the range of $\thetainit$ is restricted to the range
\begin{equation}
    \thetainit\in\lrsb{\pi/2-\theta_l,\pi/2+\theta_l},
\end{equation}
where $\cos\theta_l=\abs{l_z}$.
By marginalizing the differential rate over $L$ and $l_z$, we get the differential rate of stars as
\begin{equation}
    \frac{\partial^2\Gamma}{\partial L\partial l_z}=\frac{\pi nL}{v}.
\end{equation}
The above expression coincidentally does not depend on $l_z$.


\section{Additional Tables and Figures}\label{sec:other_tables_figures}

\begin{figure}
    \centering
    \includegraphics[width=\linewidth]{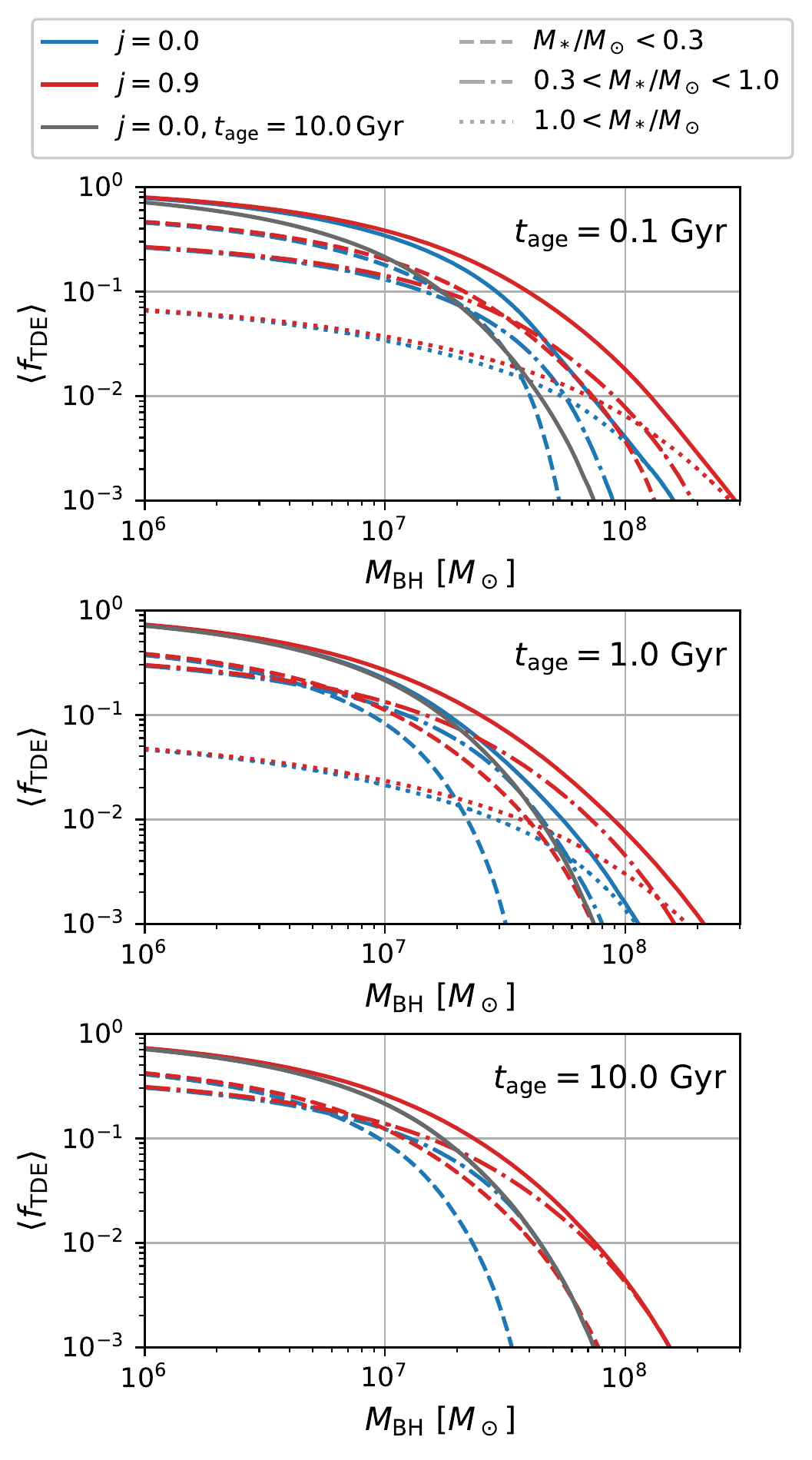}
    \caption{
    Decomposition of the observable TDE fraction $\fTDE$ into three mass ranges, $\mstar/\msun<0.3$ (dashed), $0.3<\mstar/\msun<1.0$ (dash-dotted), $1.0<\mstar/\msun$ (dotted lines). Solid lines show the stellar mass-integrated total TDE fractions.
    The blue and red lines represent different BH spins of $j=0.0,0.9$, respectively.
    To help guide the eye, the total $\fTDE$ for $j=0.0, \tage=10.0\,\Gyr$ is show in a dark gray solid line in each panel.
    }
    \label{fig:TDE_rate_decomposed_no_normalization}
\end{figure}

In this Appendix, we include Table \ref{tab:mBHmax} for the maximum BH mass for individual stars of given a mass and age and for different BH spins, Table \ref{tab:mBHmax_population} for the maximum BH mass for a Kroupa stellar population at a given age and for different BH spins, and Table \ref{tab:fTDE} for the population-averaged TDE fraction at different BH masses, spins, and stellar ages. We also include Figure \ref{fig:TDE_rate_decomposed_no_normalization} for the TDE fraction decomposed into contributions by three stellar mass bins.

\begin{table}
    \centering
    \begin{tabular}{cccc}
        \hline
        $\mstar$ & $\tage$ & $j$  & $\mBHmax$\\
        $[\msun]$ & [$\Gyr$] & & [$\msun$]\\
        \hline
           0.2 &   0.1 & 0.00 & $5.46\times10^7$ \\
           0.2 &   0.1 & 0.50 & $7.34\times10^7$ \\
           0.2 &   0.1 & 0.90 & $1.49\times10^8$ \\
           0.2 &   0.1 & 0.99 & $2.76\times10^8$ \\
           0.2 &   1.0 & 0.00 & $3.06\times10^7$ \\
           0.2 &   1.0 & 0.50 & $4.09\times10^7$ \\
           0.2 &   1.0 & 0.90 & $8.32\times10^7$ \\
           0.2 &   1.0 & 0.99 & $1.53\times10^8$ \\
           0.2 &   5.0 & 0.00 & $3.07\times10^7$ \\
           0.2 &   5.0 & 0.50 & $4.12\times10^7$ \\
           0.2 &   5.0 & 0.90 & $8.35\times10^7$ \\
           0.2 &   5.0 & 0.99 & $1.56\times10^8$ \\
           1.0 &   0.1 & 0.00 & $1.19\times10^8$ \\
           1.0 &   0.1 & 0.50 & $1.54\times10^8$ \\
           1.0 &   0.1 & 0.90 & $3.11\times10^8$ \\
           1.0 &   0.1 & 0.99 & $5.80\times10^8$ \\
           1.0 &   1.0 & 0.00 & $1.22\times10^8$ \\
           1.0 &   1.0 & 0.50 & $1.59\times10^8$ \\
           1.0 &   1.0 & 0.90 & $3.20\times10^8$ \\
           1.0 &   1.0 & 0.99 & $5.95\times10^8$ \\
           1.0 &   5.0 & 0.00 & $1.30\times10^8$ \\
           1.0 &   5.0 & 0.50 & $1.69\times10^8$ \\
           1.0 &   5.0 & 0.90 & $3.36\times10^8$ \\
           1.0 &   5.0 & 0.99 & $6.29\times10^8$ \\
           3.7 &   0.1 & 0.00 & $3.29\times10^8$ \\
           3.7 &   0.1 & 0.50 & $4.23\times10^8$ \\
           3.7 &   0.1 & 0.90 & $8.38\times10^8$ \\
           3.7 &   0.1 & 0.99 & $1.55\times10^9$ \\
        \hline
    \end{tabular}
    \caption{
    The maximum BH mass of TDE-hosting galactic nuclei $\mBHmax$, which is defined by $\fTDE(\mBHmax)=10^{-3}$, for different stellar masses $\mstar$, ages $\tage$, and BH spins $j$.
    }
    \label{tab:mBHmax}
\end{table}

\begin{table}
    \centering
    \begin{tabular}{ccc}
    \hline
    $\tage$ & $j$ & $\lara{\mBHmax}$ \\
    $[\Gyr]$ & & $[\msun]$ \\
    \hline
      0.1 & 0.00 & $1.59\times10^8$ \\
      0.1 & 0.50 & $1.76\times10^8$ \\
      0.1 & 0.90 & $2.87\times10^8$ \\
      0.1 & 0.99 & $4.72\times10^8$ \\
      1.0 & 0.00 & $1.14\times10^8$ \\
      1.0 & 0.50 & $1.29\times10^8$ \\
      1.0 & 0.90 & $2.12\times10^8$ \\
      1.0 & 0.99 & $3.51\times10^8$ \\
     10.0 & 0.00 & $7.41\times10^7$ \\
     10.0 & 0.50 & $8.61\times10^7$ \\
     10.0 & 0.90 & $1.53\times10^8$ \\
     10.0 & 0.99 & $2.59\times10^8$ \\
     \hline
    \end{tabular}
    \caption{
    The population-averaged maximum BH mass $\lara{\mBHmax}$ (eq. \ref{eq:mBHmax_population}) that can produce observable TDE for different spins $j$ and ages $\tage$. The stellar population considered have stars formed simultaneously from the Kroupa IMF.
    }
    \label{tab:mBHmax_population}
\end{table}

\begin{table}
    \centering
    \begin{tabular}{ccccc}
    \hline
    $\mBH$ & $j$ & $\tage$ & $\lara{\fTDE}$ & $\lara{\GammaTDE}$\\
    $[10^7\,\msun]$ & & $[\Gyr]$ & & $[\rm gal^{-1}yr^{-1}]$\\
    \hline
            1.00 & 0.00 &          0.1 & 0.344 & $1.36\times10^{-5}$ \\
            1.00 & 0.50 &          0.1 & 0.354 & $1.39\times10^{-5}$ \\
            1.00 & 0.90 &          0.1 & 0.384 & $1.51\times10^{-5}$ \\
            1.00 & 0.99 &          0.1 & 0.400 & $1.58\times10^{-5}$ \\
            1.00 & 0.00 &          1.0 & 0.222 & $8.77\times10^{-6}$ \\
            1.00 & 0.50 &          1.0 & 0.234 & $9.22\times10^{-6}$ \\
            1.00 & 0.90 &          1.0 & 0.269 & $1.06\times10^{-5}$ \\
            1.00 & 0.99 &          1.0 & 0.288 & $1.14\times10^{-5}$ \\
            1.00 & 0.00 &          5.0 & 0.220 & $8.67\times10^{-6}$ \\
            1.00 & 0.50 &          5.0 & 0.231 & $9.12\times10^{-6}$ \\
            1.00 & 0.90 &          5.0 & 0.266 & $1.05\times10^{-5}$ \\
            1.00 & 0.99 &          5.0 & 0.286 & $1.13\times10^{-5}$ \\
            1.00 & 0.00 &         10.0 & 0.214 & $8.45\times10^{-6}$ \\
            1.00 & 0.50 &         10.0 & 0.226 & $8.91\times10^{-6}$ \\
            1.00 & 0.90 &         10.0 & 0.261 & $1.03\times10^{-5}$ \\
            1.00 & 0.99 &         10.0 & 0.281 & $1.11\times10^{-5}$ \\
            4.97 & 0.00 &          0.1 & 0.028 & $1.11\times10^{-6}$ \\
            4.97 & 0.50 &          0.1 & 0.037 & $1.46\times10^{-6}$ \\
            4.97 & 0.90 &          0.1 & 0.070 & $2.75\times10^{-6}$ \\
            4.97 & 0.99 &          0.1 & 0.091 & $3.60\times10^{-6}$ \\
            4.97 & 0.00 &          1.0 & 0.013 & $5.02\times10^{-7}$ \\
            4.97 & 0.50 &          1.0 & 0.016 & $6.34\times10^{-7}$ \\
            4.97 & 0.90 &          1.0 & 0.033 & $1.31\times10^{-6}$ \\
            4.97 & 0.99 &          1.0 & 0.050 & $1.98\times10^{-6}$ \\
            4.97 & 0.00 &          5.0 & 0.010 & $4.01\times10^{-7}$ \\
            4.97 & 0.50 &          5.0 & 0.014 & $5.35\times10^{-7}$ \\
            4.97 & 0.90 &          5.0 & 0.031 & $1.21\times10^{-6}$ \\
            4.97 & 0.99 &          5.0 & 0.048 & $1.88\times10^{-6}$ \\
            4.97 & 0.00 &         10.0 & 0.006 & $2.56\times10^{-7}$ \\
            4.97 & 0.50 &         10.0 & 0.010 & $3.80\times10^{-7}$ \\
            4.97 & 0.90 &         10.0 & 0.026 & $1.04\times10^{-6}$ \\
            4.97 & 0.99 &         10.0 & 0.043 & $1.70\times10^{-6}$ \\
    \hline
    \end{tabular}
    \caption{
    Population-averaged TDE fraction $\lara{\fTDE}$ for different ages $\tage$ and spins $j$, at two selected BH masses, $\log_{10}(\mBH/\msun)=7.0$ and $7.7$.
    The observable TDE rate $\lara{\GammaTDE}$ is obtained from $\lara{\GammaTDE}=\lara{\fTDE}\cdot\dot{N}$, where the normalization is taken to be $\dot{N}=3.94\times10^{-5}\,\rm gal^{-1}yr^{-1}$ by setting $\log_{10}(\mBH/\msun)=7.35$ (the geometric mean value) in eq. (\ref{eq:power-law_loss_cone_rate}).
    }
    \label{tab:fTDE}
\end{table}

\label{lastpage}
\end{document}